\documentclass[letter]{aa} 

\usepackage{graphicx}
\usepackage{txfonts}
\begin{document}

\title{What makes red quasars red?}
\subtitle{Observational evidence for dust extinction from line ratio analysis}

\author{Dohyeong Kim \inst{1,2} \and Myungshin Im \inst{1,2}}

\institute{Center for the Exploration of the Origin of the Universe (CEOU),
Astronomy Program, Department of Physics and Astronomy, Seoul National University, 
1 Gwanak-ro, Gwanak-gu, Seoul 151-742, South Korea
\and 
Astronomy Program, Department of Physics and Astronomy, Seoul National University, 1 Gwanak-ro, Gwanak-gu, Seoul 151-742, South Korea\\
\email{dohyeong@astro.snu.ac.kr; mim@astro.snu.ac.kr}}

%   \institute{Center for the Exploration of the Origin of the Universe (CEOU),
%   Astronomy Program, Department of Physics and Astronomy, Seoul National University, 
%   Shillim-Dong, Kwanak-Gu, Seoul 151-742, South Korea\\
%              \email{wuchterl@amok.ast.univie.ac.at}
%         \and
%             Astronomy Program, Department of Physics and Astronomy, Seoul National University, Shillim-Dong, Kwanak-Gu, Seoul 151-742, South Korea\\
%             \email{c.ptolemy@hipparch.uheaven.space}
%             \thanks{The university of heaven temporarily does not
%                     accept e-mails}
%             }

   \date{Received September 15, 1996; accepted March 16, 1997}

% \abstract{}{}{}{}{} 
% 5 {} token are mandatory
 
\abstract
{Red quasars are very red  in the optical through near-infrared (NIR) wavelengths, which is
 possibly due to dust extinction in their host galaxies as expected
 in a scenario in which red quasars are an intermediate population between merger-driven star-forming galaxies and unobscured type 1 quasars.
 However, alternative mechanisms also exist to explain their red colors:
 (i) an intrinsically red continuum;
 (ii) an unusual high covering factor of the hot dust component, that is, $\rm CF_{HD} = {\it L}_{HD} / {\it L}_{bol}$,
 where the ${L}_{\rm HD}$ is the luminosity from the hot dust component and
 the ${L}_{\rm bol}$ is the bolometric luminosity;
 and (iii) a moderate viewing angle.
 In order to investigate why red quasars are red, we studied optical and NIR spectra of 20 red quasars at $z\sim$0.3 and 0.7,
 where the usage of the NIR spectra allowed us to look into red quasar properties in ways that are little affected by dust extinction.
 The Paschen to Balmer line ratios were derived for 13 red quasars and the values were found to
 be $\sim$10 times higher than unobscured type 1 quasars,
 suggesting a heavy dust extinction with $A_V > 2.5$ mag.
 Furthermore, the Paschen to Balmer line ratios of red quasars are difficult to explain with
 plausible physical conditions without adopting the concept of the dust extinction.
 The $\rm CF_{HD}$ of red quasars are similar to, or marginally higher than, those of unobscured type 1 quasars.
 The Eddington ratios, computed for 19 out of 20 red quasars,
 are higher than those of unobscured type 1 quasars
 (by factors of $3 \sim 5$), and hence the moderate viewing angle scenario is disfavored.
 Consequently, these results strongly suggest the dust extinction
 that is connected to an enhanced nuclear activity
 as the origin of the red color of red quasars,
 which is consistent with the merger-driven quasar evolution scenario.}

   \maketitle
%
%________________________________________________________________
\section{Introduction}

 Large area surveys in X-ray, ultraviolet (UV), optical, and radio
 wavelengths have uncovered nearly a half million quasars to date
 \citep{grazian00,becker01,anderson03,croom04,risaliti05,schneider05,veron-cetty06,im07,lee08,young09,paris14,kim15c}.
 In the UV to optical wavelength range, the spectra of quasars
 show a blue power-law continuum with broad and narrow emission lines
 or a host galaxy continuum with only narrow emission lines, which are classified as type 1 and 2 quasars, respectively.
 The unification model \citep{urry95} has been proposed to explain different types of quasars.
 In the model, a quasar is composed of
 a black hole (BH), accretion disk, dust torus, broad line regions (BLRs), and narrow line regions (NLRs).
 Under the unification model, the type 1 and 2 quasars are physically the same,
 but an obscuring dust torus prevents us from seeing the accretion disk and the BLR
 in a certain line-of-sight direction for type 2 quasars.

 However, our knowledge about quasars is still incomplete,
 especially with a population of quasars called red quasars,
 which have been identified in recent infrared surveys
 \citep{webster95,benn98,cutri01,glikman07,glikman12,glikman13,urrutia09,banerji12,stern12,assef13,fynbo13,lacy13}.
 Red quasars typically indicate quasars with red continua from UV or optical through NIR
 (e.g., $r'-K>5$ mag and $J-K>1.3$ mag in \citealt{urrutia09}; $g-r>0.5$ in \citealt{young08}),
 red mid-infrared (MIR) colors \citep{lacy04},
 or detections in X-rays while being obscured in UV and optical wavelengths
 \citep{norman02,anderson03,risaliti05,young09,brusa10}.% Depending on how they are selected, red quasars can be either type 1, type 2, or both
% (e.g., \citealt{urrutia09,brusa10}).

 Interestingly, red quasars are expected in simulations where merger-driven galaxy evolution is stressed
 \citep{menci04,hopkins05,hopkins06,hopkins08}.
 In such simulations, major mergers of galaxies trigger both star formation and quasar activity,
 often appearing as ultra-luminous infrared galaxies (ULIRGs; \citealt{sanders88,sanders96}).
 After that, the central BH grows rapidly but it is still obscured by the remaining dust in their host galaxies.
 Finally, the obscured quasar evolves to an unobscured quasar after the feedback from the central BH sweeps away cold gas and dust.
 In this picture, red quasars should appear during the intermediate stage
 between ULIRGs and unobscured quasars.
 These quasars are red owing to dust extinction by the remaining dust and gas in their host galaxies.
 So far, observational signatures of red quasars tend to support this picture.
 For example, red quasars have
 (i) higher BH accretion rates than unobscured type 1 quasars by factors of 4 to 5 \citep{urrutia12,kim15b},
 (ii) enhanced star formation activities \citep{georgakakis09},
 (iii) a high fraction of merging features \citep{urrutia08,glikman15}, and
 (iv) young radio jets \citep{georgakakis12}.

 However, several studies have proposed varying explanations for the red continuum of red quasars.
 \citet{wilkes02} and \citet{rose13} suggested that the red $J-K$ colors ($J-K>2$) of red quasars come from
 a moderate viewing angle in the unification model when the photons from
 the accretion disk and the BLR are blocked by the dust torus and not by the dust in host galaxies.
 Other studies suggest that some ($\sim$40\%; \citealt{young08}) red quasars have intrinsically red continuum
 \citep{puchnarewicz98,whiting01,young08,rose13,ruiz14}
 or an unusual covering factor of hot dust \citep{rose14}.
 Also a synchrotron emission peak at NIR wavelength from radio jet \citep{whiting01}
 has been proposed as a possible reason for the intrinsically red continuum of red quasars.

 The hydrogen line ratios can be used to infer the amount of dust extinction over the BLRs.
 \citet{rose13} showed that 2MASS selected ($J-K>2$) red quasars have higher $L_{\rm H\alpha}$/$L_{\rm H\beta}$ ratios
 than unobscured type 1 quasars. However, the difference is modest
 ($4.9\pm0.5$ versus $3.2\pm0.4$ for red and unobscured type 1 quasars, respectively)
 and the modest difference could originate from not only the dust extinction but also different physical condition in the BLRs.

 In this study, we use P$\beta$ and P$\alpha$ lines with Balmer lines to investigate the red colors of red quasars
 that are shown to have broad emission lines (i.e., type 1).
 The use of the line ratios over a wide range of wavelength makes it easier to understand
 if the red colors are due to dust extinction or other mechanisms.
 Additionally, the use of AGN NIR diagnostics (e.g., \citealt{kim10})
 allows us to measure black hole masses and bolometric luminosities in a way that is almost dust free.
 This fact is very advantageous when trying to distinguish several plausible mechanisms for red colors.
 For example, in a simple viewing angle scenario,
 we expect to find red quasars to have Eddington ratios similar to ordinary type 1 quasars,
 but not so in the intermediate population scenario.
 Throughout this paper, we use a standard $\Lambda$CDM cosmological model of
 $H_{0}$=70 km s$^{-1}$ Mpc$^{-1}$, $\Omega_{m}$=0.3, and $\Omega_{\Lambda}$=0.7,
 supported by previous observations (e.g., \citealt{im97}).

\section{Sample and data}
 In this work, we used  20 red quasars at $0.186 < z < 0.842$
 that are composed of 16 red quasars ($z>0.5$ and $L_{\rm bol} > \rm{10^{46}\,erg\,s^{-1}}$) studied in \citet{kim15b}
 and four additional red quasars ($z<0.5$ and $L_{\rm bol} \sim \rm{10^{46}\,erg\,s^{-1}}$).
 These 20 red quasars are a subsample of $\sim 80$ spectroscopically confirmed red quasars
 in \citet{glikman07} and \citet{urrutia09}, which were selected to be red quasars
 based on their broadband colors ($R-K>4$ and $J-K>1.7$ mag in \citealt{glikman07};
 $r'-K>5$ and $J-K>1.3$ mag in \citealt{urrutia09})
 from radio-detected 2MASS point sources.
 In this work, \textquotedblleft radio-detected\textquotedblright\  means the detection of the object in the FIRST radio catalog \citep{becker95}.
 We note that  radio-detected AGNs are not necessarily radio-loud.
 We define  radio-loudness as $R_{i}=\log (f({\rm 1.4\,GHz})/f({\rm 7480\,\AA{}}))$ \citep{ivezic02}.
 In our red quasars sample, we find that only one-third are radio-loud (e.g., $R_{i}>2$; \citealt{karouzos14}) and the remaining two-thirds are in the radio-intermediate regime ($1<R_{i}<2$).
 Additionally, these quasars are chosen to be at $z \sim 0.3$ or $z \sim 0.7$ so that
 we can sample both the Balmer and Paschen lines.

\begin{figure}
 \centering
% \figurenum{1}
 \includegraphics[scale=0.4]{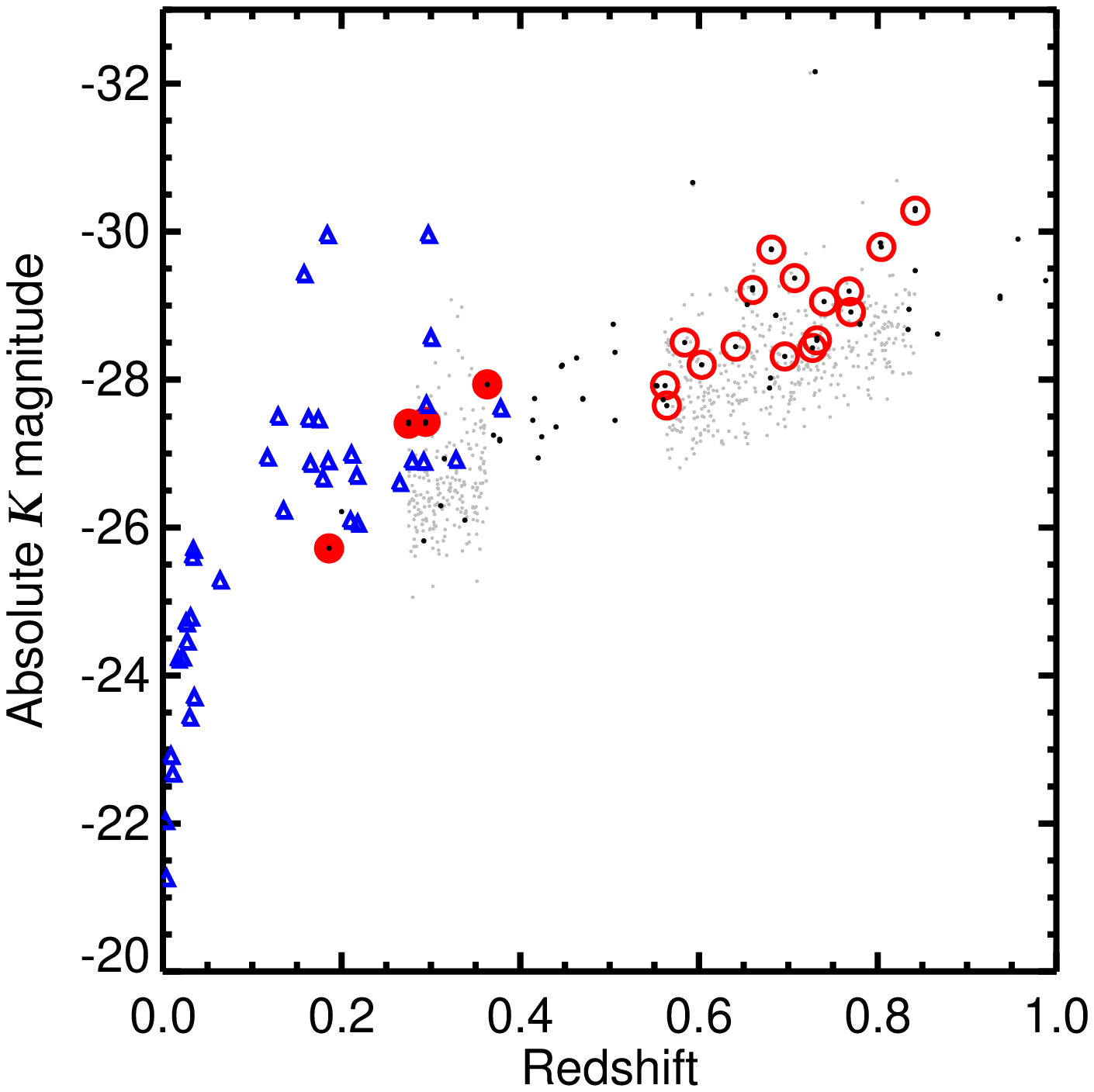}\\
 \caption{Redshifts and absolute $K$-band magnitudes of red quasars and unobscured type 1 quasars.
 The 20 red quasars used in this study are denoted by red open circles,
 and the black dots indicate the red quasars listed in \citet{glikman07} and \citet{urrutia09}
 from which our sample is drawn.
 Four red quasars (0036$-$0113, 0817$+$4354, 1209$-$0107, and 1307$+$2338),
 which were not used in \cite{kim15b}, are denoted by filled red circles.
% Two red quasars (0036$-$0113 and 1307$+$2338) are marked by filled red circles, of which the spectra are provided in this paper.
 The comparison sample is made of 37 unobscured type 1 quasars from Kim et al. (2010; blue open triangles), and 623 SDSS unobscured type 1 quasars at $0.275 < z < 0.363$ and $0.562 < z < 0.842$ (gray dots).}
 \label{Fig.1}
\end{figure}

 In this study, we compare several properties of red quasars to those of unobscured type 1 quasars.
 We used 623 unobscured type 1 quasars at $z \sim$ 0.3 and 0.7,
 both selected from Sloan Digital Sky Survey (SDSS) in an approach similar to the red quasars (also see Section 5.3),
 and 37 unobscured type 1 quasars at $z < 0.5$ for which Paschen line information is available from \cite{kim10}.
 The 37 unobscured type 1 quasars are relatively bright ($K < $\,14.5 mag).
 The unobscured type 1 quasars cover a wide range in the $K$ band luminosities ($-30.0 < K\,{\rm mag} < -21.3$),
 which overlaps well with the $K$-band luminosities of our red quasars ($-30.3 < K\,{\rm mag} < -25.7$).
  Although the unobscured type 1 quasars from \cite{kim10} include six faint ($K\,{\rm mag} > -24$) sources,
 the host galaxy contamination is negligible ($< 8$\,\%) due to a narrow slit width \citep{glikman06,kim10}.

 Figure 1 indicates the redshifts and absolute $K$-band magnitudes of the 20 red quasars and the comparison quasar sample.
 The absolute $K$-band magnitudes are $K-$corrected
 using a composite spectrum of unobscured type 1 quasars (NIR: \citealt{glikman06}; MIR: \citealt{kim15a}).
 Here, we do not apply the dust extinction correction,
 and we note that the amount of extinction is rather small for $K$ band.
 The mean $E(B-V)$ of red quasars is 0.88 ($A_K =0.45$ mag).
% (e.g., 0.7 mag for $E(B-V)=2.0$ mag).

 For 13 of the 20 red quasars, both the optical and NIR spectra are available
 and the remaining objects have only NIR spectra.
 The optical spectra of the red quasars come from \citet{glikman07,glikman12}
 while the NIR spectra are from \citet{glikman07,glikman12} and \citet{kim15b}.
 Additionally, new NIR spectra were obtained for two red quasars (0036$-$0113 and 1307$+$2338) using
 the SpeX instrument \citep{rayner03} on the NASA Infrared Telescope Facility (IRTF).
 The observation was performed with a set of the short cross-dispersion mode (SXD; 0.8--2.5\,$\mu$m)
 and a $0 \farcs 8$ slit width to achieve a resolution of $R \sim 750$ (400\,$\rm km~s^{-1}$ in FWHM)
 under clear weather and good seeing conditions of $\sim 0\farcs 7$.
 In order to get fully reduced spectra, we used the Spextool software \citep{vacca03,cushing04}.
 The spectrophotometric calibration was carried out using the standard star spectra taken before or after
 the spectrum of each quasar was obtained.
 These flux-calibrated spectra were matched with a $K-$band photometry
 from 2MASS and additional flux scaling was performed when necessary.
 The fully reduced and calibrated NIR spectra of the two red quasars are provided in Table 3 by a machine readable table form.

 Among our sample, we detected P$\beta$ in 19 and P$\alpha$ in two red quasars
 (only one red quasar is detected in both P$\beta$ and P$\alpha$ lines) at signal-to-noise ratios (S/N) of $>5$.
 Among 13 red quasars with optical spectra, we detected H$\beta$ in 11 and H$\alpha$ in 3 red quasars at S/N of $>5$.
 For 2 red quasars (0817$+$4354 and 1656$+$3821), we derived upper limits of their H$\beta$ line luminosities.
 For the remainder, the spectra at H$\beta$ and H$\alpha$ lines are not readily available.

 For the lines with no detection, we set upper limits.
 For the upper limits of the H$\beta$ line luminosity,
 we assume the FWHM of typical red quasars of 3000\,$\rm km\,s^{-1}$ \citep{kim15b}.
 Then, the 5-sigma limit of the integrated luminosity density within the wavelength width of the FWHM value is adopted.

 As summarized in Table 1, 12 red quasars have both H$\beta$ and P$\beta$ detections or upper limits,
 while only P$\beta$ line is detected for 7 red quasars.
 A few red quasars have H$\alpha$ or P$\alpha$ line detected.

\section{Analysis}
 We corrected the spectra for the Galactic extinction by adopting the reddening map of \cite{schlafly11},
 and the extinction curve with $R_V=3.1$ of \citet{fitzpatrick99}.
 Then, we shifted the corrected spectra to the rest frame.
 In order to estimate the H$\beta$ luminosity, first we determined the continuum spectrum around the H$\beta$ line.
 Spectrum of type 1 quasar around the H$\beta$ line is generally complex,
 which contains a power-law continuum component, a host galaxy component,
 blended \ion{Fe}{II} multiplets, and several narrow emission lines
 (e.g., [\ion{O}{II}] $\lambda$3726, H$\gamma$ $\lambda$4340, and [\ion{O}{III}] $\lambda \lambda$4959, 5007 doublet).

\begin{figure}
 \centering
% \figurenum{2}
 \includegraphics[width=\columnwidth]{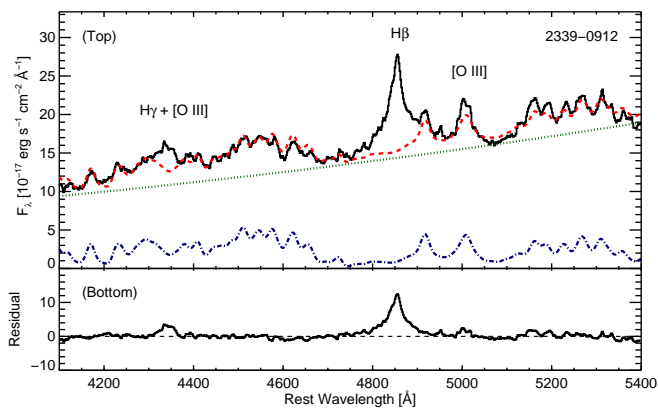}\\ %[scale=0.30]
 \caption{(Top) Optical spectrum of 2339$-$0912 (the black solid line) and its model spectrum components.
 The optical spectrum includes several emission lines such as H$\gamma$, H$\beta$, and [\ion{O}{III}].
 The red dashed line indicates the fitted continuum spectrum
 that is composed of a power-law component (green dotted line) and a component for the Fe blends (blue dot-dashed line).
 (Bottom) The continuum-subtracted spectrum showing broad Balmer emission lines.
 The continuum-subtracted spectra are used for fitting the lines.}
 \label{Fig.2}
\end{figure}

 We fit the continuum of the red quasars with the power-law component and Fe template \citep{boroson92}.
 To avoid contamination from the several narrow emission lines around the H$\beta$ line, the continuum-fitting regions were chosen as
 4150--4280\,\AA, 4400--4750\,\AA, 4910--4945\,\AA, and 5070--5400\,$\rm \AA{}$.
 For the fit, a \texttt{MPFIT} \citep{markwardt09} code based on an interactive data language (IDL) procedure was used to
 determine the power-law continuum and Fe blends.
 The Fe blends were determined by broadening and scaling the Fe template from the spectrum of IZw1 \citep{boroson92}.
 Figure 2 shows the optical spectrum of 2339$-$0912 with the fitted continuum model
 (the power-law continuum and  Fe blends) and the continuum-subtracted spectrum.

 We omitted the host galaxy component from the fit for two reasons.
 First, among the 11 Balmer line measurable red quasars,
 the measured rest-frame 5100\,$\rm \AA{}$ continuum luminosities ($L_{\rm 5100} =  \lambda L_{\lambda}$ at 5100\,\AA)
 of 6 red quasars (0825$+$4716, 1113$+$1244, 1227$+$5053, 1434$+$0935, 1720$+$6156, and 2339$-$0912)
 are very luminous (extinction-corrected $L_{\rm 5100} = L_{\rm bol} / 9.2 > \rm{10^{45}\,erg,s^{-1}}$; Table 2).
 For type 1 AGNs as luminous as the red quasars in $L_{\rm 5100}$ (after the extinction correction),
 the host galaxy contamination is known to be well below 1\% (\citealt{shen11}),
 and we assume the same for the red quasars.
 Then, even if we assume that the dust extinction occurs only in the nuclear part
 \footnote{If the dust extinction obscures the light from both the nucleus and host galaxy by the same amount,
 then the host galaxy contamination stays at $<1\,\%$.} with $E(B-V)=1$,
 the host contamination would be $<30\,\%$ at 5100\,$\rm \AA{}$.
 Second, we measured the equivalent width (EW) of \ion{Ca}{II} K $\lambda$3934 absorption line for the remaining 5 red quasars,
 since an EW of 1.5\,$\rm \AA{}$ for \ion{Ca}{II} K corresponds to the host galaxy contribution of $\sim$10\,\% \citep{greene05,kim06}.
 We could not detect the \ion{Ca}{II} K absorption line for 3 red quasars,
 and for the remaining 2, we measured EWs of 3.8\,$\rm \AA{}$, which
 corresponds to the host galaxy contribution of $\sim$25\,\% (0036$-$0113 and 1307$+$2338),
 by scaling the results of \cite{greene05} and \cite{kim06}.
 Even if the host galaxy contamination of red quasars is bigger than 25\,\%,
 we stress that it will not affect the line luminosity and width measurements much since the continuum is well subtracted for each object.

\begin{figure}
 \centering
% \figurenum{3}
 \includegraphics[width=\columnwidth]{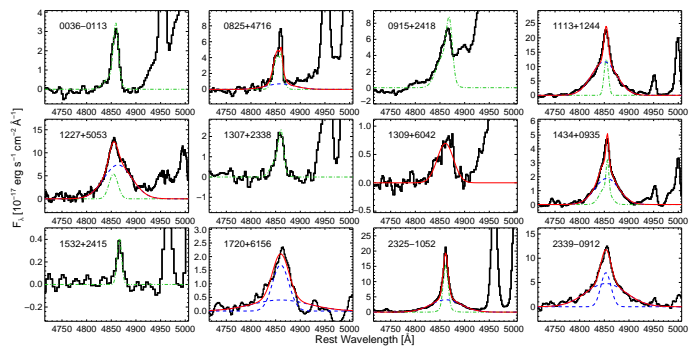}\\
 \caption{Results of the fitting of the H$\beta$ lines. The continuum is already subtracted.
 The black and red solid lines indicate the observed spectra and best-fit model, respectively.
 The narrow component of H$\beta$ line is fitted by using the nearby [\ion{O}{III}] line at 5007\,$\rm \AA{}$ as a template,
 and the fitted narrow component is represented by the green dash-dotted line.
 The dashed blue lines represent the broad component ($\rm FWHM > 800$\,$\rm km^{-1}$) of the H$\beta$ line.}
 \label{Fig.3}
\end{figure}

 After the continuum subtraction, we fit the narrow component of H$\beta$ line
 using [\ion{O}{III}] line at 5007\,$\rm \AA{}$ as a template when possible.
 The [\ion{O}{III}] line was fitted with a double Gaussian function
 to reproduce both the symmetric component and asymmetric outflow component \citep{greene05},
 and the fitted result was used as the template.
 This was possible for nine red quasars.

 Among the nine [\ion{O}{III}]-fitted red quasars, the H$\beta$ line of four red quasars
 (0036$-$0113, 0915$+$2418, 1307$+$2338, and 1532$+$2415)
 is well fitted by the narrow component only,
 although 0036$-$0113 and 1307$+$2338 were reported to have a broad H$\beta$ component \citep{glikman07}.
 For these four red quasars, we estimated the upper limit of the H$\beta$ line luminosities.
 In order to estimate the upper limit of the H$\beta$ line luminosities,
 we performed the same procedures (see Section 2) in the narrow component subtracted spectra.

 For the remaining five [\ion{O}{III}]-fitted red quasars (0825$+$4716, 1113$+$1244, 1227$+$5053, 1434$+$0935, and 2325$-$1052),
 the H$\beta$ line is fitted by a composition of the broad and narrow component.
 For the fit, we set the central wavelength, line width, and line flux of the broad component as free parameters.
 We set the lower limit of FWHM of broad component as 800\,$\rm km\,s^{-1}$
 that is twice of the typical FWHM of narrow emission line (400\,$\rm km\,s^{-1}$; \citealt{schneider06})
 and the FWHMs of the fitted broad component are bigger than 800\,$\rm km\,s^{-1}$.

\begin{figure}
 \centering
% \figurenum{4}
 \includegraphics[width=\columnwidth]{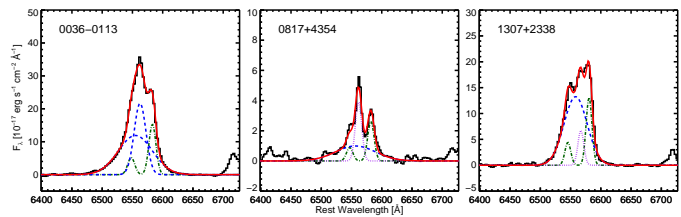}\\
 \caption{Results of the fitting of the H$\alpha$ lines.
 The black and red solid lines indicate the continuum subtracted spectra and best-fit model, respectively.
 The blue dashed and purple dotted lines are the best-fit models for the broad and narrow component of the H$\alpha$ line, respectively,
 and the green dot-dashed lines indicate the best-fit models for the [\ion{N}{II}] doublet.}
 \label{Fig.4}
\end{figure}

 For the [\ion{O}{III}]-unfitted three red quasars (1309$+$6042, 1720$+$6156, and 2339$-$0912),
 we fit broad H$\beta$ line with a single (1309$+$6042) or double (1702$+$6156 and 2339$-$0912) Gaussian function,
 and the fit is performed by the \texttt{MPFIT}.
 Similar to the procedure above, the central wavelength, line width, and line flux are set to be free parameters.
 The FWHMs of the fitted components of the three red quasars are bigger than 800\,$\rm km\,s^{-1}$
 and this result means that the fitted component represent the broad components.
 The flux of the H$\beta$ line is taken to be the sum of the two broad components for 1704$+$6156 and 2339$-$0912,
 while only the flux of the single broad component is taken to be the broad line flux for 1309$+$6042.
 Figure 3 shows the H$\beta$ line-fitting results.

 To fit the H$\alpha$ line, we used a similar procedure as for H$\beta$ line fitting.
 However, for the fitting the continuum, we use a single power-law component only because of
 the negligible contribution of the Fe blends around the H$\alpha$ line.
 After the continuum subtraction, fitting the H$\alpha$ is somewhat more complicated than fitting the H$\beta$ line
 because the H$\alpha$ line is blended with a [\ion{N}{II}] $\lambda \lambda$ 6548, 6583 doublet.
 For H$\alpha$ line measurable red quasars 0036$-$0113, 0817$+$4354, and 1307$+$2338,
 we constructed a model for narrow component of H$\alpha$ and [\ion{N}{II}] doublet
 using nearby [\ion{S}{II}] $\lambda \lambda$ 6716, 6731 doublet.
 Using the model of narrow component, we simultaneously fit the H$\alpha$ line and [\ion{N}{II}] doublet.
 The [\ion{N}{II}] doublet is also fitted by two single Gaussian functions,
 but the central wavelength and flux ratio (2.96; \citealt{kim06}) are fixed.
 For two red quasars (0817$+$4354 and 1307$+$2338),
 the H$\alpha$ line is fitted by a composition of broad and narrow component,
 and the narrow component is from the model of the [\ion{S}{II}] doublet.
 For the other red quasar (0036-0113),
 the H$\alpha$ line is fitted by double broad components.
 Only the flux of the broad component is taken to be the broad H$\alpha$ flux for 0817$+$4354 and 1307$+$2338,
 but the sum of the fluxes from the two broad components is taken to be the broad component flux of H$\alpha$ line for 0036$-$0113.
 We show the fitting results of the H$\alpha$ lines in Figure 4.

 The P$\beta$ and P$\alpha$ lines are detected for 19 and 2 red quasars, respectively.
 Among these, the P$\beta$ luminosities of 16 red quasars are adopted from \citet{kim15b}.
 The Paschen line luminosities of the other red quasars are newly measured, following the procedure in \citet{kim10}.
 Figure 5 shows the fitting of the Paschen lines.

 This procedure fits the line with a single or double Gaussian functions, neglecting the NLR component,
 since the S/N and the resolution of the spectra do not allow us to measure the NLR component.
 The measured line luminosities and line widths are taken as the values for the broad line.
 Hence, the derived broad line luminosities and widths can be somewhat biased.
 To correct for the bias we applied the mean correction factors that are derived from
 well-resolved Balmer lines of 26 local unobscured type 1 quasars \citep{kim10}.
 They measured the fluxes and FWHMs of the well-resolved Balmer lines with a single, double, and multiple Gaussian functions,
 and derived the correction factors by comparing the measured properties.
 The mean correction factors are $\rm flux_{multi}$/$\rm flux_{double} = 1.05$
 and $\rm flux_{multi}$/$\rm flux_{single} = 1.06$ for correcting the luminosities and
 $\rm FWHM_{multi}$/$\rm FWHM_{double} = 0.85$ and $\rm FWHM_{multi}$/$\rm FWHM_{single} = 0.91$ for correcting FWHM values.
 For more details, see Section 2.3 of \cite{kim10}.

\begin{figure}
 \centering
% \figurenum{5}
 \includegraphics[width=\columnwidth]{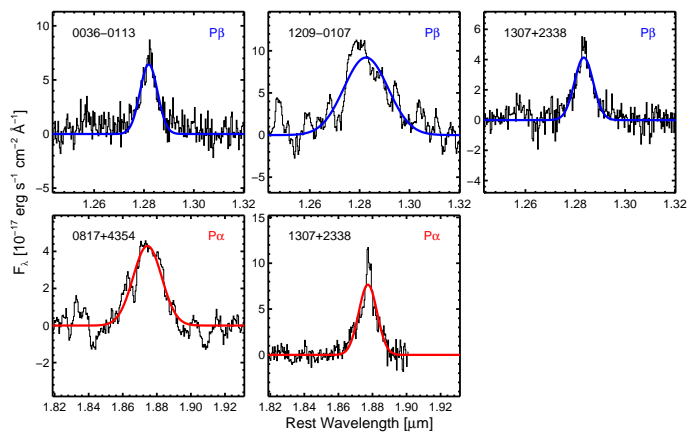}\\
 \caption{Fitting results of the P$\beta$ and P$\alpha$ lines.
 The black solid lines are the continuum subtracted spectra.
 The blue and red solid lines are the best-fit models for the P$\beta$ and P$\alpha$ lines, respectively.}
 \label{Fig.5}
\end{figure}

 When no H$\beta$ line is detected (0817$+$4354 and 1656$+$3821),
 we estimated the upper limit of the H$\beta$ line luminosity.
 In order to estimate the upper limit of the H$\beta$ line luminosity, we used the same procedure mentioned above.
% First, we measure the standard deviation of the continuum subtracted fluxes in
% a wavelength range of 48.61\,$\rm \AA{}$ away from the H$\beta$ line,
% where the width of 48.61\,$\rm \AA{}$ corresponds to 3000\,$\rm km\,s^{-1}$,
% which is similar to FWHM of a broad line of a typical red quasar (e.g., \citealt{kim15b}).
% Then, we assume the single Gaussian profile that has a FWHM of 3000\,$\rm km\,s^{-1}$
% and has a peak three times bigger than the standard deviation.
 The measured upper limits of H$\beta$ line luminosities of the two red quasars are provided in Table 1.

\begin{sidewaystable*}
%\begin{table*}
\centering
\caption{Luminosities and FWHMs of BLRs for red quasars \label{tbl1}}
\begin{tabular}{cccccccccccccccccccccccc}
\hline\hline
\noalign{\smallskip}
Object Name&                            Redshift&                       &
$\log L_{\rm H\beta}$           &       $\rm FWHM_{H\beta}$&    &
$\log L_{\rm H\alpha}$& $\rm FWHM_{H\alpha}$&   &
$\log L_{\rm P\beta}$&  $\rm FWHM_{P\beta}$&    &
$\log L_{\rm P\alpha}$& $\rm FWHM_{P\alpha}$\\
        &                                               ($z$)&                          &
($\rm{erg~s^{-1}}$)&            ($\rm km\,s^{-1}$)&     &
($\rm{erg~s^{-1}}$)&            ($\rm km\,s^{-1}$)&     &
($\rm{erg~s^{-1}}$)&            ($\rm km\,s^{-1}$)&     &
($\rm{erg~s^{-1}}$)&            ($\rm km\,s^{-1}$)\\
\noalign{\smallskip}
\hline
\noalign{\smallskip}
0036$-$0113&    0.294&  &       $<$40.9&                --&                             &       42.58$\pm$0.04& 1476$\pm$135&   &       42.22$\pm$0.02& 1765$\pm$73&    &       --&                             --\\
0817$+$4354&    0.186&  &       $<$40.7&                --&                             &       41.02$\pm$0.11& 3022$\pm$452&   &       --&                             --&                             &       41.97$\pm$0.04& 3172$\pm$270\\
0825$+$4716&    0.804&  &       42.21$\pm$0.17& 4199$\pm$1541&  &       --&                             --&                             &       43.62$\pm$0.04& 3940$\pm$393&   &       --&                             --\\
0911$+$0143&    0.603&  &       --&                             --&                             &       --&                             --&                             &       42.74$\pm$0.04& 2907$\pm$206&   &       --&                             --\\
0915$+$2418&    0.842&  &       $<$43.1&                --&                             &       --&                             --&                             &       43.77$\pm$0.09& 3527$\pm$729&   &       --&                             --\\
1113$+$1244&    0.681&  &       43.20$\pm$0.06& 3699$\pm$350&   &       --&                             --&                             &       43.28$\pm$0.03& 2054$\pm$107&   &       --&                             --\\
1209$-$0107&    0.363&  &       --&                             --&                             &       --&                             --&                             &       42.99$\pm$0.02& 4451$\pm$246&   &       --&                             --\\
1227$+$5053&    0.768&  &       43.18$\pm$0.06& 4160$\pm$409&   &       --&                             --&                             &       42.98$\pm$0.08& 2781$\pm$446&   &       --&                             --\\
1248$+$0531&    0.752&  &       --&                             --&                             &       --&                             --&                             &       42.95$\pm$0.02& 2697$\pm$123&   &       --&                             --\\
1307$+$2338&    0.275&  &       $<$40.8&                --&                     &       42.17$\pm$0.05& 2050$\pm$69&    &       41.99$\pm$0.02& 1891$\pm$107&   &       42.41$\pm$0.03& 2029$\pm$124\\
1309$+$6042&    0.641&  &       41.70$\pm$0.07& 2275$\pm$245&   &       --&                             --&                             &       42.71$\pm$0.02& 2673$\pm$89&    &       --&                             --\\
1313$+$1453&    0.584&  &       --&                             --&                             &       --&                             --&                             &       43.08$\pm$0.01& 3793$\pm$96&    &       --&                             --\\
1434$+$0935&    0.770&  &       42.52$\pm$0.08& 3538$\pm$387&   &       --&                             --&                             &       42.99$\pm$0.01& 1398$\pm$34&    &       --&                             --\\
1532$+$2415&    0.562&  &       $<$40.9&                --&                     &       --&                             --&                             &       42.88$\pm$0.16& 4071$\pm$1549&  &       --&                             --\\
1540$+$4923&    0.696&  &       --&                             --&                             &       --&                             --&                             &       42.88$\pm$0.14& 5397$\pm$1518&  &       --&                             --\\
1600$+$3522&    0.707&  &       --&                             --&                             &       --&                             --&                             &       43.31$\pm$0.06& 1887$\pm$228&   &       --&                             --\\
1656$+$3821&    0.732&  &       $<$43.3&                --&                             &       --&                             --&                             &       43.15$\pm$0.04& 3028$\pm$246&   &       --&                             --\\
1720$+$6156&    0.727&  &       42.51$\pm$0.16& 2866$\pm$305&   &       --&                             --&                             &       42.70$\pm$0.06& 1818$\pm$205&   &       --&                             --\\
2325$-$1052&    0.564&  &       42.63$\pm$0.04& 4522$\pm$2871&  &       --&                             --&                             &       42.90$\pm$0.13& 2778$\pm$772&   &       --&                             --\\
2339$-$0912&    0.660&  &       43.12$\pm$0.01& 2191$\pm$196&   &       --&                             --&                             &       43.29$\pm$0.03& 2927$\pm$140&   &       --&                             --\\
\noalign{\smallskip}
\hline
\end{tabular}
%\tablefoot{\\
%$\ast$: The H$\beta$ line is fitted by a narrow component only.
%}
%\tablecomments{$\ast$: The $E(B-V)_{\rm cont}$ values are adopted from \cite{glikman07},
%which uncertainties are not available.}
%\end{table*}
\end{sidewaystable*}

 In total, we obtained the broad line luminosities of eight H$\beta$,
 three H$\alpha$, 19 P$\beta$, and two P$\alpha$ lines.
 The line luminosities and FWHMs are summarized in Table 1.
 We note that all the values in Table 1 are only corrected for Galactic extinction.

\section{Results}
 In order to investigate the factor leading to the red color of red quasars,
 we compared the luminosity ratios of the P$\beta$/H$\beta$, P$\beta$/H$\alpha$, P$\alpha$/H$\beta$, and P$\alpha$/H$\alpha$
 of red quasars to those of unobscured type 1 quasars.
 For a typical red quasar with the $E(B-V)=2$ mag,
 its H$\alpha$ and H$\beta$ luminosities would be suppressed by a factor of 100 and 1000, respectively,
 but the P$\alpha$ and P$\beta$ fluxes are suppressed by a factor of only 2.3 and 4.7, respectively.
 Therefore, the analysis using optical to NIR emission lines is a useful way to investigate the dust obscuration,
 where the Paschen lines can serve as a good measure of the unobscured light.

\begin{figure*}
 \centering
% \figurenum{6}
 \includegraphics[scale=0.43]{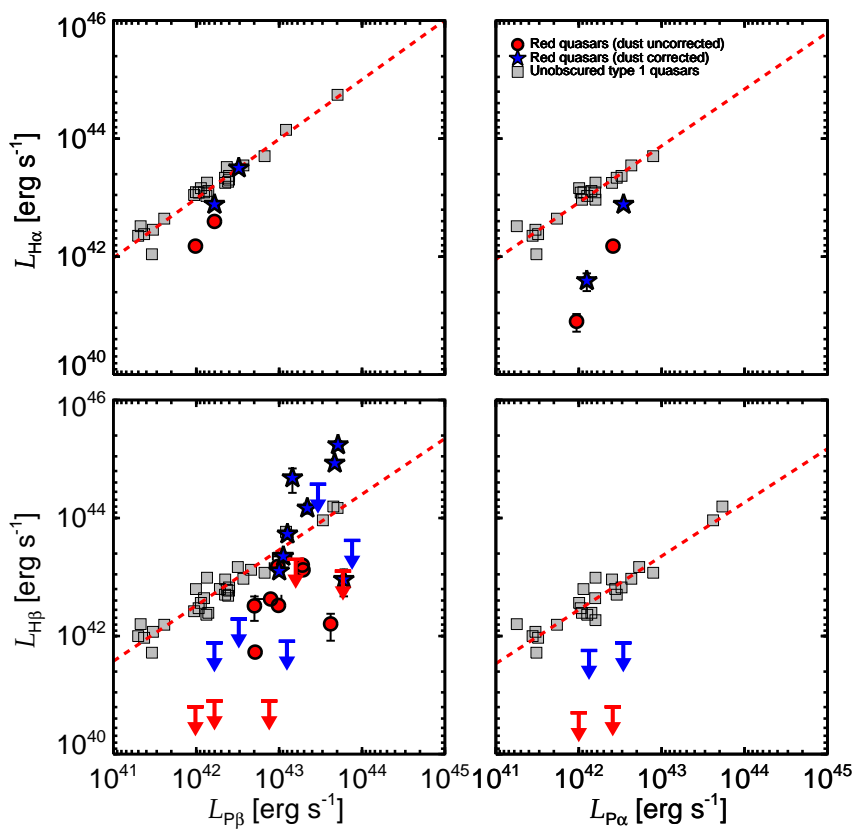}\\ %[width=\textwidth]
 \caption{Comparison between the Paschen and Balmer line luminosities of red quasars and unobscured type 1 quasars (gray squares).
 The dashed red lines mean the best-fit lines between the Paschen and Balmer line luminosities of unobscured type 1 quasars.
 The blue stars and red circles indicate the dust-corrected and uncorrected luminosities of the red quasars, respectively.
 The blue and red arrows represent the upper limits of dust corrected and uncorrected $L_{\rm H\beta}$ vs. $L_{\rm P\beta}$
 of 0036$-$0113, 0915$+$2418, 1307$+$2338, 1532$+$2415, and 1656$+$3821, respectively,
 and $L_{\rm H\beta}$ vs. $L_{\rm P\alpha}$ of 0817$+$4354 and 1307$+$2338, respectively.
 The left two panels show the comparison of the P$\beta$ and Balmer line luminosities,
 and the right two panels compare the P$\alpha$ with the Balmer line luminosities.
 Before extinction correction and at a given Paschen line luminosity,
 the Balmer line luminosities of red quasars are much weaker than those of unobscured type 1 quasars,
 while extinction corrections bring the red quasars close to the unobscured type 1 quasars.
 }
 \label{Fig.6}
\end{figure*}

 For this analysis, the line luminosity ratios of unobscured type 1 quasars are adopted from \citet{kim10}.
 Although several faint ($L_{\rm Paschen} < 10^{42}\,{\rm erg\,s^{-1}}$) unobscured type 1 quasars
 are included in this comparison, the line luminosity ratios are insensitive to the luminosity
 (see Figure 6 in \citealt{kim10}).
 Figure 6 compares line luminosities of the red quasars and unobscured type 1 quasars.
 We find that the observed Balmer line luminosity for a given Paschen line luminosity is
 1.5--290 ($\sim$12 on average) times weaker for the red quasars than for unobscured type 1 quasars.

\begin{figure}[h]
 \centering
% \figurenum{7}
 \includegraphics[scale=0.35]{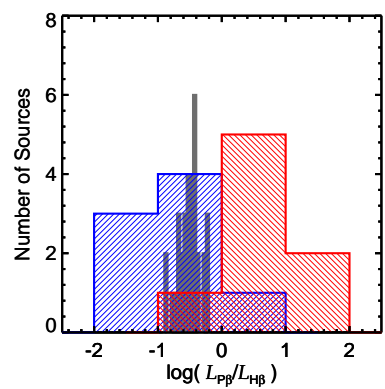}\\
 \caption{Distributions of the P$\beta$/H$\beta$ luminosity ratios of the red quasars (color-hatched histogram) and unobscured type 1 quasars (gray histogram).
 The blue and red histograms indicate the distributions of the extinction corrected and uncorrected
 P$\beta$/H$\beta$ luminosity ratios of the red quasars, respectively.
 After applying the extinction correction, the P$\beta$/H$\beta$ distribution of red quasars agree broadly with that of unobscured type 1 quasars,
 but with a much larger scatter suggesting that the extinction-correction prescription is not perfect.
 }
 \label{Fig.7}
\end{figure}

 Figure 7 shows the distributions of the P$\beta$/H$\beta$ luminosity ratios of the red quasars and unobscured type 1 quasars.
 We find that red quasars have the $\log ( L_{\rm P\beta} / L_{\rm H\beta})$ values much higher than
 those of the unobscured type 1 quasars; median $\log ( L_{\rm P\beta} / L_{\rm H\beta})$ of the red quasars are 0.27$\pm$0.53,
 in contrast to that of unobscured type 1 quasars (-0.49$\pm$0.17).
 The Kolmogorov-Smirnov (K-S) test confirms this significant difference in the line luminosity ratios between the two quasar populations.
 We performed the K-S test using the \texttt{KSTWO} code based on the IDL.
 The maximum deviation between the cumulative distributions of these two P$\beta$/H$\beta$ luminosity ratios, D, is 1.0,
 and the probability of the result given the null hypothesis, $p$, is only 9.22$\times \rm{10^{-7}}$.

\begin{figure*}[t]
 \centering
% \figurenum{8}
 \includegraphics[scale=0.55]{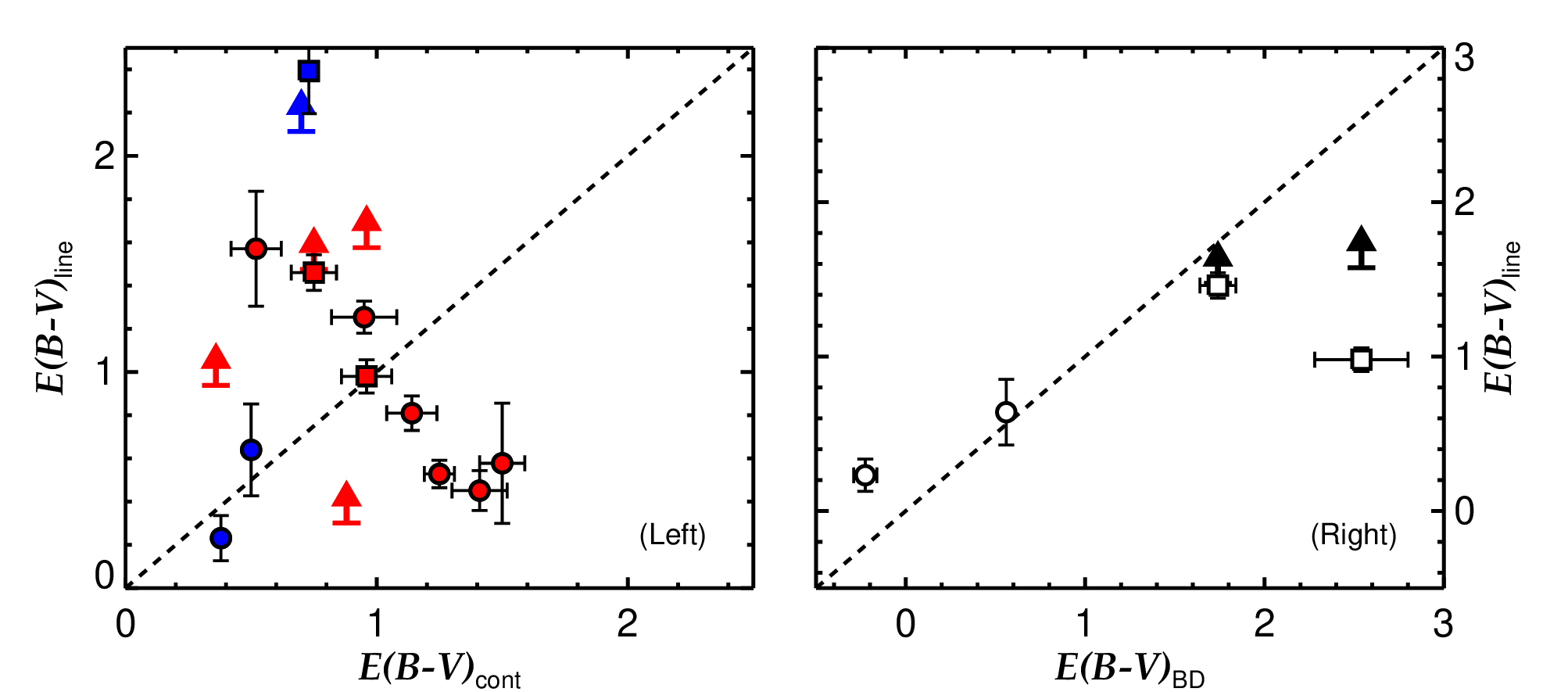}\\
 \caption{(Left) Comparison between the continuum-based and the line ratio-based $E(B-V)$ values of red quasars.
 The $E(B-V)_{\rm line}$ values are composed of $E(B-V)_{\rm line 1}$ and $E(B-V)_{\rm line 2}$ values.
 The $E(B-V)_{\rm line 1}$ and $E(B-V)_{\rm line 2}$ are represented by the circles and squares, respectively.
 The $E(B-V)_{\rm cont}$ values are adopted from previous studies \citep{glikman07,urrutia09},
 for which uncertainties for some of the values are not available.
 The red and blue colors indicating the $E(B-V)_{\rm cont}$ values are adopted from \cite{urrutia09} and \cite{glikman07}, respectively.
 The arrows representing the $E(B-V)_{\rm line 1}$ values are measured using the upper limits luminosities of H$\beta$ and $L_{\rm P\beta}$,
 and their colors have the same meanings.
 The dashed line indicates a line where the two values are identical.
 The correlation between these two quantities are nonexistent,
 suggesting large uncertainties that are associated with the determination of
 the dust extinction from the continuum that may be affected by the modeling of the continuum components.
 (Right) The $E(B-V)_{\rm BD}$ vs. $E(B-V)_{\rm line}$.
 The meaning of the circles, square, and arrows are identical to the left panel.
 The $E(B-V)_{\rm BD}$ values are adopted from \cite{glikman07},
 and they agree well with the $E(B-V)_{\rm line}$.}
 \label{Fig.8}
\end{figure*}

 If red quasars are dust-reddened,
 we expect the dust-corrected line luminosity ratios of the red quasars to
 be consistent with those of unobscured type 1 quasars.
 For this test, we adopted the $E(B-V)$ values from previous studies \citep{glikman07,urrutia09}
 and applied the extinction correction assuming the Galactic extinction curve with $R_V=3.1$ \citep{schultz75}.
 \cite{glikman07} and \cite{urrutia09} provide two types of $E(B-V)$ values derived from the continuum shape and the Balmer decrement.
 In this study, we used the $E(B-V)$ values from the continuum shape because
 the $E(B-V)$ values from the continuum shape are available for all objects in the sample.
 The extinction-corrected line luminosities ratios of red quasars are indicated in Figures 6 and 7.

 The extinction-corrected median $\log ( L_{\rm P\beta} / L_{\rm H\beta})$ of the red quasars is -0.63$\pm$0.81,
 that is almost the same as that of unobscured type 1 quasars.
 Furthermore, our K-S test for the histogram of the extinction-corrected P$\beta$/H$\beta$ luminosity ratios
 show that the measured D and $p$ values of the K-S statistic are 0.44 and 0.12, respectively, against unobscured type 1 quasars.
 For this K-S test, we broadened the P$\beta$/H$\beta$ luminosity ratios distribution
 of unobscured type 1 quasars by adding scatters to P$\beta$/H$\beta$ luminosity ratios through
 a Monte Carlo simulation by the amount that could be produced during dereddening of
 P$\beta$/H$\beta$ luminosity ratios of red quasars assuming a typical scatter in $E(B-V)$ of 0.5 \citep{glikman07}.
 Therefore, we conclude that, on average, both the continuum colors and the line ratios of red quasars
 can be explained by dust extinction.

 One interesting question is whether we can accurately determine $E(B-V)$ values of red quasars.
 In an attempt to make this determination, we compared $E(B-V)$ values from the Paschen and Balmer line luminosity ratios ($E(B-V)_{\rm line}$)
 versus $E(B-V)$ values from the continuum shape ($E(B-V)_{\rm cont}$).
 In this study, the $E(B-V)_{\rm line}$ is composed of  the two values $E(B-V)_{\rm line 1}$ and $E(B-V)_{\rm line 2}$.
 The $E(B-V)_{\rm line 1}$ is derived by finding a proper value of $E(B-V)$ that
 brings the observed $L_{\rm P\beta}$/$L_{\rm H\beta}$ ratio to the observed mean of unobscured type 1 quasars, where
 the Galactic extinction curve and $R_V=3.1$ are assumed.
 On the other hand, the $E(B-V)_{\rm line 2}$ is determined by using the ratios of
 $L_{\rm P\beta}$ (0817$+$4354 and 1307$+$2338) or $L_{\rm P\alpha}$ (0036$-$0113) to $L_{\rm H\alpha}$.
 The measured $E(B-V)_{\rm line 1}$ and $E(B-V)_{\rm line 2}$ values are listed in Table 2.
 In Figure 8, we compare $E(B-V)_{\rm line}$ versus $E(B-V)_{\rm cont}$.
 The Pearson correlation coefficient between the two quantities is -0.21 and a rms scatter of 0.68 with respect to a one-to-one correlation.
 The result can be considered as no correlation or a one-to-one correlation with a very large scatter of 0.72 in $E(B-V)$.
 On the same plot, we also compare $E(B-V)_{\rm line}$ versus $E(B-V)$ from the Balmer decrement ($E(B-V)_{\rm BD}$).
 The $E(B-V)_{\rm BD}$ values are taken from \cite{glikman07} and determined using the ratios between the $L_{\rm H\alpha}$ and $L_{\rm H\beta}$.
 We note that the $E(B-V)_{\rm BD}$ values were determined using
 total line intensity or broad component luminosity.
 Among the two types of $E(B-V)_{\rm BD}$, we use the $E(B-V)_{\rm BD}$ from the broad component luminosity
 except for 1227$+$5053 for which $E(B-V)_{\rm BD}$ from the broad line is not available.
 For the comparison between the $E(B-V)_{\rm BD}$ and $E(B-V)_{\rm line}$,
 only four data points are available, but we find a reasonably good agreement between
  $E(B-V)_{\rm line}$ and $E(B-V)_{\rm BD}$.
 The Pearson correlation coefficient between the two quantities is 0.80.

 The large scatter between $E(B-V)_{\rm cont}$ and $E(B-V)_{\rm line}$ is likely due to the wide range of continuum slopes that quasars can have
 and the difficulty in estimating the intrinsic continuum shape in advance.
 Another possible reason is that the dust obscuration region varies between the continuum emitting and line emitting regions.
 We suggest that $E(B-V)$ estimated through the continuum shape contains a large scatter ($\sim 0.72$ in $E(B-V)$).
 This agrees with what we saw in Figure 7, where the histogram of $L_{\rm P\beta}$/$L_{\rm H\beta}$ after the extinction correction
 with $E(B-V)_{\rm cont}$ is much broader than the same histogram for unobscured type 1 quasars
 and the two histograms are virtually indistinguishable after taking into account this broadening effect.

\section{Discussion}
\subsection{Physical condition as a cause for high line luminosity ratio}
\begin{figure}[t]
 \centering
% \figurenum{9}
 \includegraphics[width=\columnwidth]{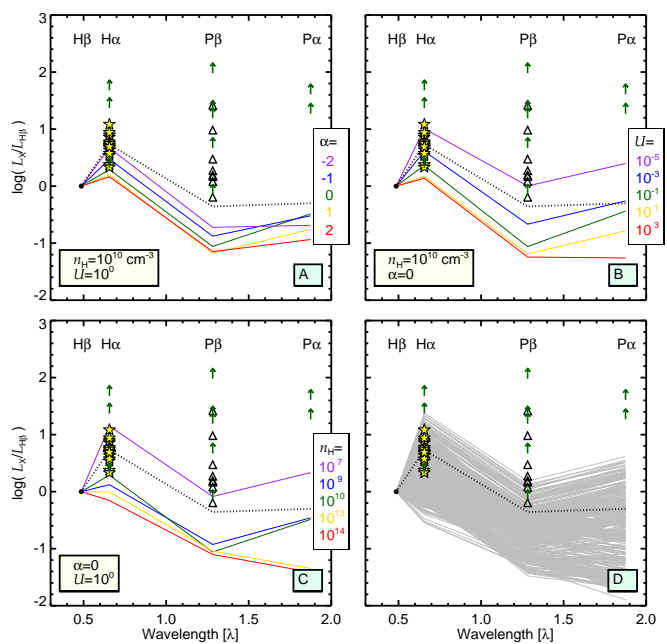}\\
 \caption{      %\scriptsize
 Line luminosity ratios of H$\alpha$, P$\beta$, and P$\alpha$ with respect to H$\beta$ of red quasars.
 The P$\beta$/H$\beta$ luminosity ratios of eight red quasars are indicated by the open triangles, and
 the yellow stars are the H$\alpha$/H$\beta$ luminosity ratios of the local red AGNs from Rose et al. (2013).
 The green arrows indicate the lower limits.
 The dotted line is the line luminosity ratio reproduced by the \texttt{CLOUDY} code with a set of parameters
 ($\alpha= -1.0$, $U=10^{-1.5}$, and $n_{\rm H}=\rm{10^{9}~cm^{-3}}$),
 which is consistent with the line luminosity ratios of unobscured type 1 quasars from H$\beta$ to P$\alpha$ \citep{kim10}.
 (A) The five colored solid lines represent the line luminosity ratios from the \texttt{CLOUDY} code
 with $n_{\rm H}={\rm 10^{10}\,cm^{-3}}$, $U=10^{0}$, and various $\alpha$ from -2 to 2.
 The various $\alpha$ is represented by color, i.e.,
 the purple, blue, green, yellow, and red lines indicate $\alpha$ is -2, -1, 0, 1, and 2, respectively.
 (B) The meanings of the symbols and lines are identical to (A),
 but the five colored lines are the line luminosity ratios with $n_{\rm H}={\rm 10^{10}\,cm^{-3}}$, $\alpha=0$,
 and various $U$ from $10^{-5}$ to $10^{3}$.
 (C) The meanings of the symbols and lines are identical to (A),
 but the five colored lines are the line luminosity ratios with $\alpha=0$, $U=10^{0}$,
 and various $n_{\rm H}$ from $10^{7}$ to $\rm 10^{14}\,cm^{-3}$.
 (D) The meanings of the symbols and the dotted line are identical to (A),
 and the gray lines represent the line luminosity ratios computed from the \texttt{CLOUDY} code
 with various physical conditions, within $\alpha = -2$ to 2, $n_{\rm H}=\rm{10^{7} - 10^{14}~cm^{-3}}$, and $U={\rm 10^{-5} - 10^{5}}$.}
 \label{Fig.9}
\end{figure}

 Here, we test a hypothesis that
 the observed $L_{\rm Paschen}$/$L_{\rm Balmer}$ of red quasars are due to
 a different physical condition of broad emission line regions (BELRs) without dust extinction.
 To do so, we explore what physical conditions of BELRs can reproduce the observed line luminosity ratios
 by computing theoretically expected line luminosity ratios under
 different physical conditions using the \texttt{CLOUDY} code (version 13.03; \citealt{ferland98}).

 We set the plausible ranges of input parameters as the following.
 Quasars do not show any broad forbidden lines
 (e.g., see \citealt{glikman07,urrutia09}).
 The absence of the broad forbidden lines implies that the hydrogen density ($n_{\rm H}$)
 in the BELR is higher than the critical density of the forbidden lines,
 which gives us the lower limit of $n_{\rm{H}} = \rm{10^{7}~cm^{-3}}$.
 The upper limit of $n_{\rm H}$ is set to $\rm{10^{14}~cm^{-3}}$
 considering the existence of strong Fe blends
 ($\rm \sim 10^{12}~cm^{-3}$; \citealt{collin-souffrin82,rees89}).
 For the other parameters, we vary the values of the shape of the ionizing continuum ($\alpha = -2 \sim 2$)
 and the ionization parameter ($U= \rm{10^{-5} \sim 10^{5}}$) to cover various physical conditions.

 Line luminosity ratios of $L_{\rm P\alpha}$/$L_{\rm H\beta}$, $L_{\rm P\beta}$/$L_{\rm H\beta}$,
 and $L_{\rm H\alpha}$/$L_{\rm H\beta}$ are sensitive to $n_{\rm H}$ and $U$.
 The line luminosity ratios decrease when the $n_{\rm H}$ and $U$ values are increased,
 which is indicated in the B and C panels of Figure 9.

 Figure 9 shows the line luminosity ratios as a function of wavelength
 for red quasars and those from the \texttt{CLOUDY} calculation.
 The line luminosity ratios of unobscured type 1 quasars can be successfully reproduced by the \texttt{CLOUDY} calculation
 with a set of parameters, $\alpha= -1.0$, $U=10^{-1.5}$, and $n_{\rm H}=\rm{10^{9}~cm^{-3}}$ \citep{kim10},
 which is represented by the dotted line in Figure 9.
 For the Balmer lines, although the H$\alpha$/H$\beta$ luminosity ratios of red quasars are not measured in this study,
 these luminosity ratios of local red AGNs are only moderately different from those of unobscured type 1 quasars \citep{rose13}.

 However, the line luminosity ratios start to demand unusual physical conditions when Paschen lines are included.
 The median $\log (L_{\rm P\beta}/L_{\rm H\beta})$ is $0.43 \pm 0.53$ for red quasars,
 while it is only $-0.48 \pm 0.17$ for unobscured type 1 quasars.
 Among the eight $L_{\rm P\beta}$/$L_{\rm H\beta}$ ratios measured red quasars,
 five (63\,\%) red quasars have higher line ratios than the maximum line luminosity ratios in the \texttt{CLOUDY} calculation
 ($\log (L_{\rm P\beta}/L_{\rm H\beta}) \le 0.18$).
 In other words, the $L_{\rm P\beta}$/$L_{\rm H\beta}$ ratios of the five red quasars are much higher than
 the line luminosity ratios from all the plausible physical parameters for BLR.
 For the remaining three (38\,\%) red quasars, the $L_{\rm P\beta}$/$L_{\rm H\beta}$ ratios
 can be explained with a condition of low $n_{\rm H}={\rm 10^{7}\,cm^{-3}}$, low $U=\rm{10^{-3}}$, and $\alpha=2$.
 The parameters are found by minimizing $\chi^2$, which is a function of the line luminosity ratios
\begin{equation}
\chi^{2} = \sum_{i=1}^N \frac{(R_{{\rm observed},i} -R_{{\rm model},i})^{2}}{\sigma_{i}^{2}},
\end{equation}
 where $N$ is the number of line luminosity ratios, and two types of $R_{i}$ are the line luminosity
 ratios either from observation or the CLOUDY model, and $\sigma_{i}$ is the uncertainty in the measured line luminosity ratio.
 However, this physical condition, that is, low $n_{\rm H}$ and low $U$, is somewhat similar to that of NLRs
 (e.g., $n_{\rm e}={\rm 10^{3.5} \sim 10^{7.5}\,cm^{-3}}$ and $U \sim {\rm 10^{-2}}$;
 \citealt{osterbrock91,netzer13}), rather than the broad lines that we are studying here (FWHM\,$\rm > 800\,km\,s^{-1}$),
 and this disfavors the physical condition as a reason for the observed line luminosity ratios.

\subsection{High hot dust covering factor as a cause for redness}
 Rose et al. (2013) suggested that some of AGNs with red $J-K$ colors are red
 owing to a relatively large hot dust covering factor
 (${\rm CF_{HD}}=L_{\rm HD}/L_{\rm bol}$; \citealt{maiolino07,kim15a}).
 We address this point here with our sample.

\begin{figure*}
 \centering
% \figurenum{10}
 \includegraphics[scale=0.4]{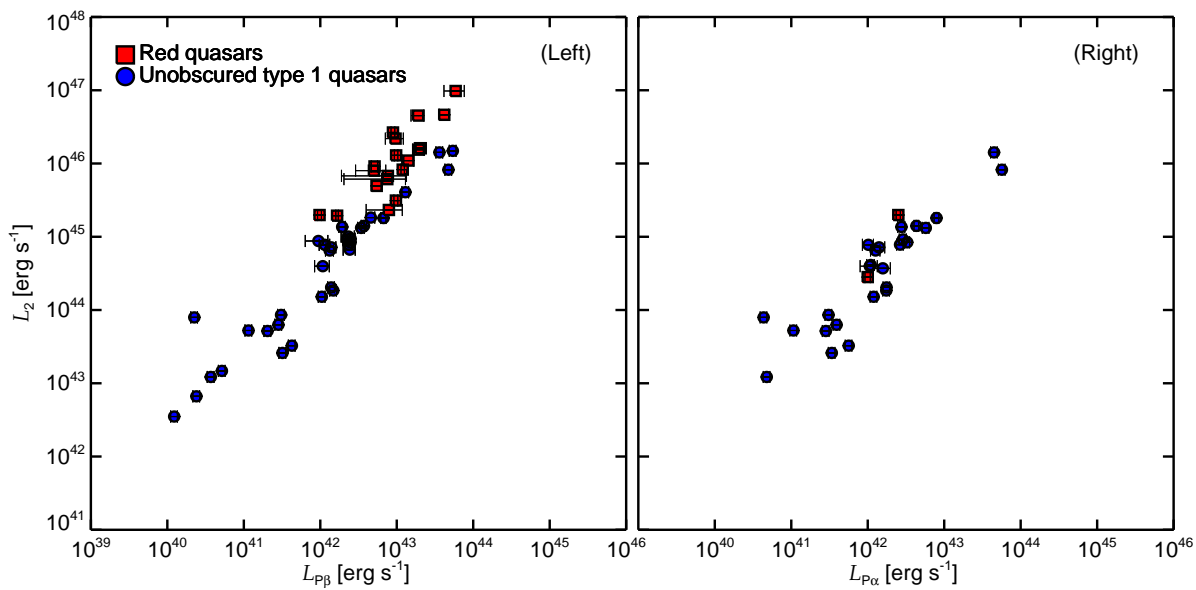}\\
 \caption{(Left) Values of $L_{\rm 2}$  vs.  $L_{\rm P\beta}$ of
 red quasars (the red filled squares) and unobscured type 1 quasars (the blue filled circles).
 (Right) The $L_{\rm 2}$ values vs. the $L_{\rm P\alpha}$ values.
 The meanings of the symbols are identical.
 This figure shows that red quasars do not have unusual hot dust emission that is much stronger ($>1$ dex)
 than that of unobscured type 1 quasars, although their NIR emission may be a bit
 stronger ($\sim 0.3$ dex) than unobscured type 1 quasars.
 }
 \label{Fig.10}
\end{figure*}

\begin{figure*}
 \centering
% \figurenum{11}
 \includegraphics[scale=0.4]{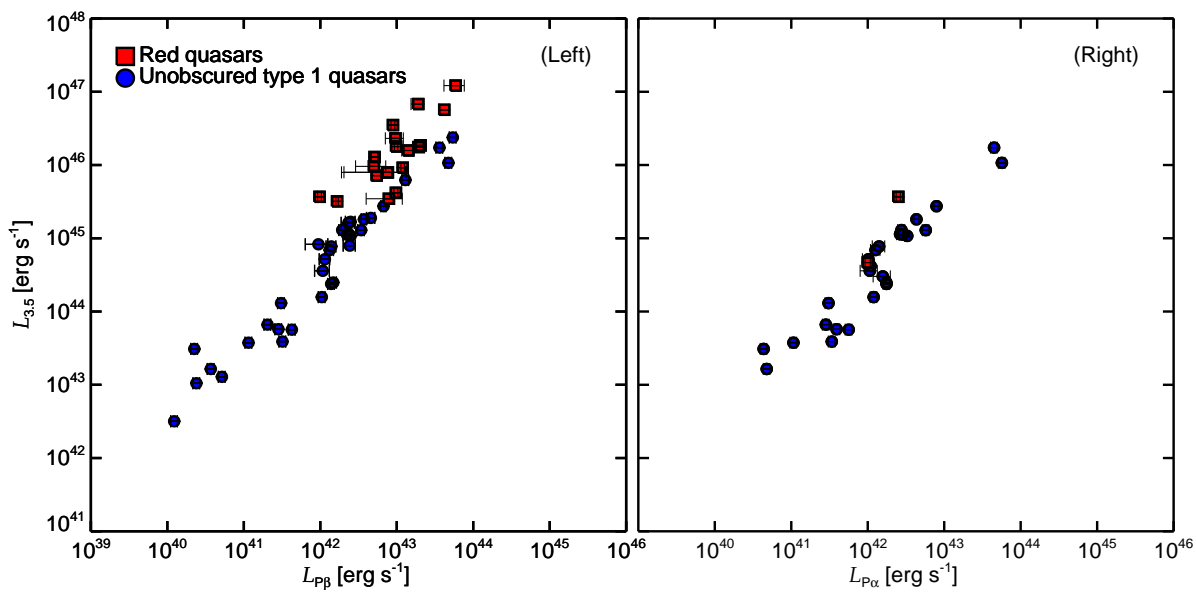}\\
 \caption{Comparisons between $L_{\rm 3.5}$ values vs. Paschen line luminosities.
 The meanings of the symbols are identical to Figure 10.}
 \label{Fig.11}
\end{figure*}

 For the comparison between the $\rm CF_{HD}$ of red quasars and unobscured type 1 quasars,
 we used 37 unobscured type 1 quasars from \citet{kim10}.
 The unobscured type 1 quasars are bright ($K<14.5$\,mag and $M_{i}<-23$\,mag) and located at $z<0.5$.

 Because the $L_{\rm bol}$ and $L_{\rm HD}$ are easily under- or overestimated from the spectral energy distribution (SED) model fitting
 depending on a set of the dust extinction value and extinction law,
 we used the Paschen line luminosities and NIR continuum luminosities
 ($L_{\rm 2}$ and $L_{\rm 3.5}$; $\lambda L_{\lambda}$ at 2\,$\mu$m and 3.5\,$\mu$m in the rest-frame)
 as proxies for the $L_{\rm bol}$ and $L_{\rm HD}$, respectively.  Considering previous results that found, first, the temperature of hot dust torus is $\sim$1000--1500\,K \citep{barvainis87,glikman06,kim15a,hernan-caballero16}
 and, second, 2\,$\mu$m and 3.5\,$\mu$m are closed to the peak wavelengths of blackbody radiation from the hot dust component,
$L_{\rm 2}$ and $L_{\rm 3.5}$ can represent $L_{\rm HD}$ \citep{glikman06,kim15a}.
 Although the stellar emission can peak at 1.6\,$\mu$m,
 the hot dust component can be a dominant component with its peak at 2--3.5\,$\mu$m.
 This is supported by $J-K$ colors of red quasars that are significantly different from
 those of normal galaxies and star-forming galaxies (e.g., see Figure 1 in \citealt{glikman12}).
 Although the NIR continuum luminosities of red quasars have a possibility to be overestimated by the stellar emission contamination,
 the NIR contribution of the stellar emission is known to be $\leq$10\,\% for quasars,
 when NIR continuum luminosity is over than $\rm 10^{43.5}\,erg\,s^{-1}$ \citep{hernan-caballero16}.
 There is no plausible reason to believe that
 the hot and warm dust emission should be much weaker for red quasars than type 1 quasars.
 If red quasars are red due to unusually high $\rm CF_{HD}$,
 this would make the host galaxy contribution to the NIR continuum luminosities even smaller.
 The Paschen line luminosities are used as a tracer for the $L_{\rm bol}$ \citep{kim15b}
 by employing the excellent correlation between $L_{\rm P}$, $L_{\rm H\alpha}$, $L_{\rm 5100}$, and $L_{\rm bol}$ \citep{kim10,shen11,jun15}.
 The wavelengths of 2\,$\mu$m and 3.5\,$\mu$m are not too far from the Paschen lines (P$\beta$: 1.2818\,$\mu$m and P$\alpha$: 1.8751\,$\mu$m)
 and so the $L_{\rm 2}$/$L_{\rm P}$ and $L_{\rm 3.5}$/$L_{\rm P}$
 are rather insensitive to the exact values of dust extinction.

 The NIR continuum luminosities are measured by interpolating the 2MASS \citep{skrutskie06} and $\it WISE$ \citep{wright10} photometric data.
 The P$\beta$ and P$\alpha$ luminosities are adopted from \cite{kim10} for 37 and 27 unobscured type 1 quasars, respectively.

 In this study, we do not consider the Baldwin effect because the Baldwin effect in the Balmer lines is weak \citep{dietrich02},
 and the Paschen lines have a strong correlation with the Balmer lines \citep{kim10}.

 Figures 10 and 11 show the comparisons between the NIR continuum luminosities versus Paschen line luminosities
 of the red quasars and unobscured type 1 quasars.
 The unobscured type 1 quasars have the mean $\log$($L_{\rm 2}$/$L_{\rm P\beta}$) and $\log$($L_{\rm 3.5}$/$L_{\rm P\beta}$)
 of 2.58$\pm$0.01 and 2.64$\pm$0.01 with dispersions of 0.55 and 0.51, respectively.
 The measured luminosity ratios are only slightly smaller than those of red quasars
 (3.06$\pm$0.01 and 3.21$\pm$0.01 with dispersions of 0.27 and 0.28).
 Moreover, the mean $\log$($L_{\rm 2}$/$L_{\rm P\alpha}$) and $\log$($L_{\rm 3.5}$/$L_{\rm P\alpha}$) of the unobscured type 1 quasars
 are 2.51$\pm$0.01 and 2.56$\pm$0.01 with dispersions of 0.64 and 0.60, respectively,
 and those of red quasars (2.72$\pm$0.02 and 2.97$\pm$0.02 with dispersions of 0.32 and 0.35) are almost same.

 The result indicates that the covering factor of red quasars are not much different from type 1 quasars,
 and an unusually large covering factor is not the reason for their red colors.
 This result is consistent with the viewing angle or torus obscuration scenario, but we address this scenario below.

\subsection{Viewing angle as a cause for redness}
 \citet{wilkes02} suggested that the redness of red quasars arises from a moderate viewing angle in the quasar unification model
 when the accretion disk and the BLR are viewed through a dust torus.
 If so, red quasars should show properties very similar to unobscured type 1 quasars.

 However, previous studies \citep{urrutia12,kim15b} showed that red quasars have significantly higher accretion rates than unobscured type 1 quasars,
 which cannot be explained by the viewing angle scenario.
 Among the 19 P$\beta$ luminosity measured red quasars in our sample,
 the Eddington ratios of 16 red quasars at $z \sim 0.7$ were already studied in \citet{kim15b}.
 We have three additional P$\beta$ measured red quasars at $z\sim 0.3$ (0036$-$0113, 1209$-$0107, and 1307$+$2338).
 Below, we first examine the Eddington ratios of the three additional quasars
 and then compare the Eddington ratios of 16 red quasars at $z \sim 0.7$ to the $L_{\rm bol}$ matched unobscured type 1 quasars.

\begin{figure}
 \centering
% \figurenum{12}
 \includegraphics[width=\columnwidth]{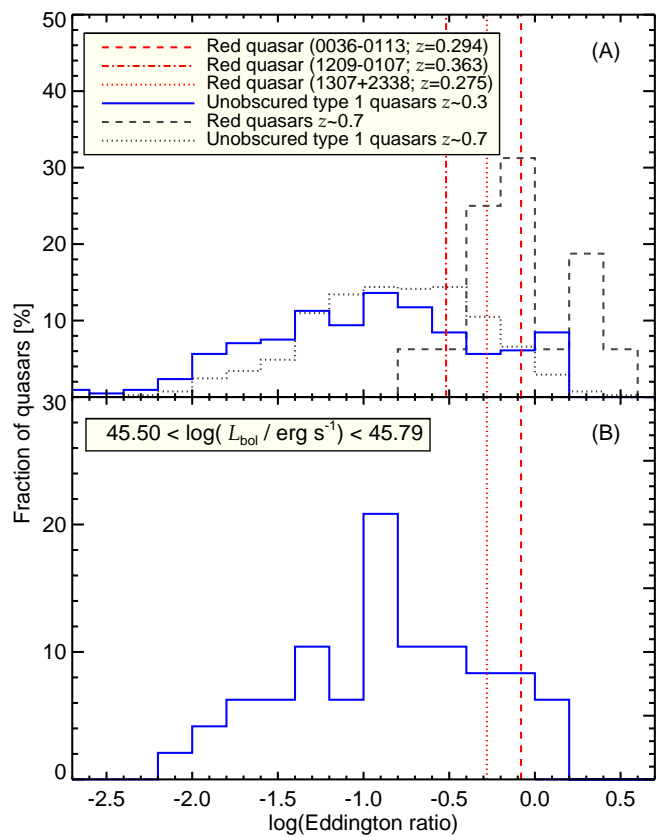}\\
 \caption{(A) Eddington ratios of three red quasars (0036$-$0113, 1209$-$0107, and 1307$+$2338; red lines)
 and unobscured type 1 quasars $z \sim 0.3$ (blue histograms).
 The gray dashed and dotted lines represent the Eddington ratio distributions for red quasars and unobscured type 1 quasars
 at $z \sim 0.7$, respectively, where results are adopted from \cite{kim15b}.
 (B) The same figure as above, except that unobscured type 1 quasars
 have bolometric luminosities similar to the two red quasars (0036$-$0113 and 1307$+$2338; $45.50 < \log (L_{\rm bol} / {\rm erg\,s^{-1}} ) < 45.79$).
 }
 \label{Fig.12}
\end{figure}

\begin{figure}
 \centering
% \figurenum{13}
 \includegraphics[width=\columnwidth]{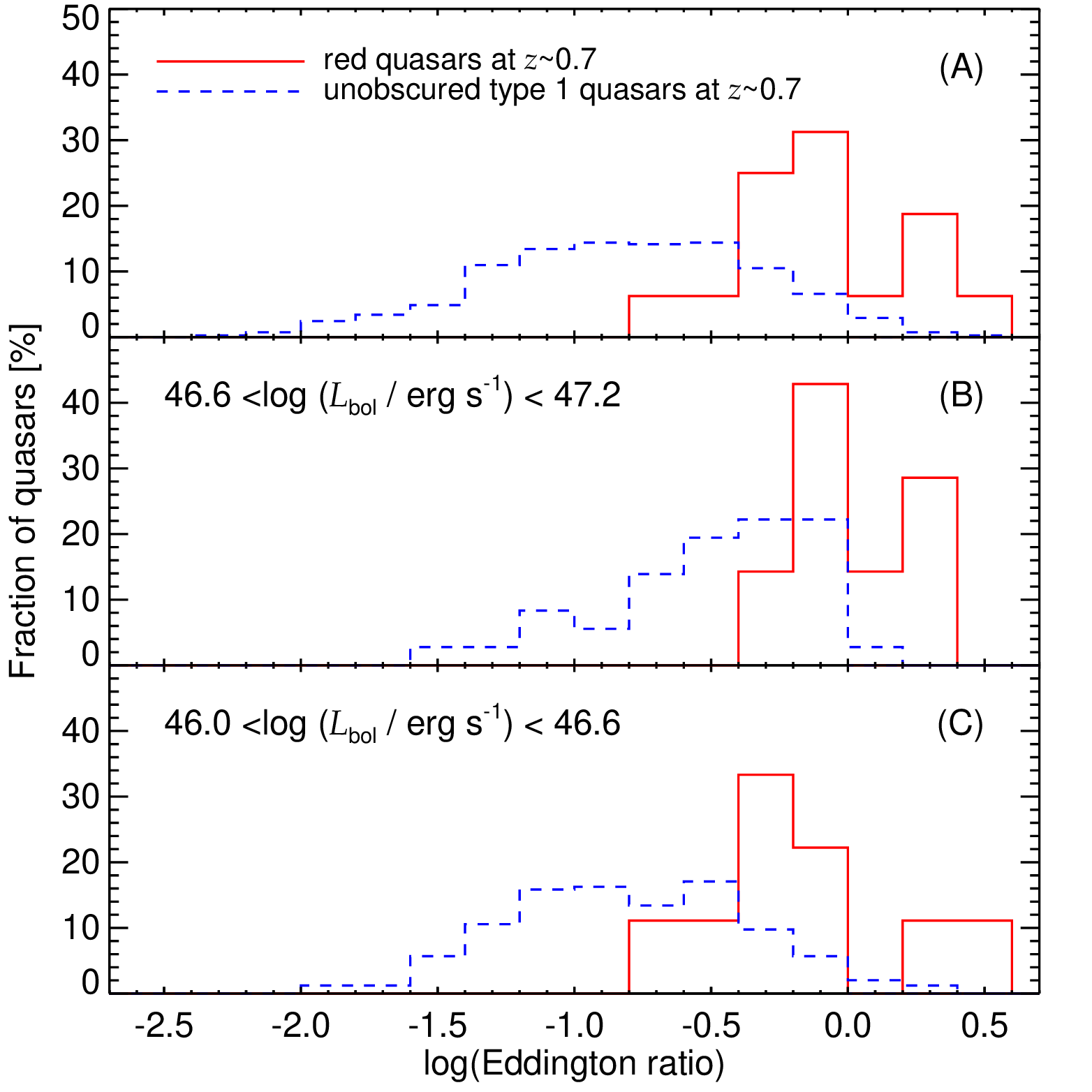}\\
 \caption{(A) Eddington ratio distributions of red quasars and unobscured type 1 quasars at $z \sim 0.7$.
 The red solid and blue dashed histograms represent red quasars and unobscured type 1 quasars, respectively.
 The Eddington ratios of red quasars at $z \sim 0.7$ are adopted from \cite{kim15b}.
 (B) Eddington ratio distributions for high-luminous quasars ($46.6 < \log ( L_{\rm bol}/{\rm erg\,s^{-1}} ) < 47.2$).
 (C) Eddington ratio distributions for low-luminous quasars ($46.0 < \log ( L_{\rm bol}/{\rm erg\,s^{-1}} ) < 46.6$).
 }
 \label{Fig.13}
\end{figure}

 As a comparison sample, we select unobscured type 1 quasars from the quasar catalog \citep{schneider10}
 of the SDSS Seventh Data Release (DR7; \citealt{abazajian09}).
 In order to avoid the sample selection bias, the unobscured type 1 quasars are selected by the same selection criteria as red quasars as follows:
 (i) the same redshift range of the three red quasars (0.275$\textless z \textless$0.363),
 (ii) radio detection in Faint Images of the Radio Sky at Twenty-Centimeters (FIRST) survey \citep{becker95},
and (iii) photometric detection in 2MASS.
 This selection yields 213 unobscured type 1 quasars as the comparison sample.

 In order to estimate the BH masses of red quasars,
 we corrected the P$\beta$ luminosities by adopting $E(B-V)$ values from previous studies \citep{glikman07,urrutia09}.
 The $E(B-V)$ values were determined using their continuum shape.
 After then, we used the P$\beta$ based $M_{\rm BH}$ estimator from Equation (1) of \citet{kim15b},
 which is modified from the P$\beta$ scaling relation \citep{kim10}
 by adopting a recent virial coefficient of $\log f = 0.05$ \citep{woo15}.
 For the $M_{\rm BH}$ of unobscured type 1 quasars,
 we used $L_{\rm 5100}$ and $\rm{FWHM_{H\beta}}$ values from \citet{shen11}
 and the $M_{\rm BH}$ estimator of Equation (2) in \citet{kim15b}.

 The H$\beta$-based $M_{\rm BH}$ values are lower than the P$\beta$-based $M_{\rm BH}$ values by 0.45 dex before the extinction correction.
 No offset can be seen after the H$\beta$-based $M_{\rm BH}$ values are derived by taking the extinction effect into account,
 but this introduces a large scatter (rms$\sim 0.56$ dex) in $M_{\rm BH}$ values from the two methods,
 justifying the use of P$\beta$-based $M_{\rm BH}$ here.

 To obtain $L_{\rm bol}$ values of the red quasars, we translated the $L_{\rm P\beta}$ using
 the relationship between the $L_{\rm bol}$ and $L_{\rm P\beta}$ of Equation (4) in \citet{kim15b}.
 The $L_{\rm bol}$ values of the unobscured type 1 quasars are converted from
 the $L_{\rm 5100}$ values with a bolometric correction factor of 9.26 \citep{shen11}.
 We note that the measured $M_{\rm BH}$ and $L_{\rm bol}$ values of the red quasars are summarized in Table 2.

 The $L_{\rm bol}$ values of the red quasars ($10^{45.50}\,{\rm erg\,s^{-1}}\,<\,{L_{\rm bol}}\,<\,10^{46.51}\,{\rm erg\,s^{-1}}$)
 and the comparison sample of unobscured type 1 quasars ($10^{45.04}\,{\rm erg\,s^{-1}}\,<\,{L_{\rm bol}}\,<\,10^{46.55}\,{\rm erg\,s^{-1}}$)
 are similar but not identical.
 For that reason, we also performed the comparison of the two populations using $L_{\rm bol}$-matched samples too.

 To estimate the Eddington ratio, the P$\beta$ is used as both an indicator of $L_{\rm bol}$ and $M_{\rm BH}$.
 Although these two different quantities are derived from the P$\beta$ line and include the same line luminosity term,
 the $M_{\rm BH}$ has a larger uncertainty than $L_{\rm bol}$ because
 the $M_{\rm BH}$ values include FWHM square term and the square-rooted line luminosity term.
 The median uncertainties of the P$\beta$ based $L_{\rm bol}$ and $M_{\rm BH}$ are 0.03 and 0.08 dex, respectively.
 The Eddington ratio is proportional to $L^{0.5}$/$\rm FWHM^2$ and this gives the combined uncertainty of 0.08 dex\footnote{The
 uncertainties only consider the uncertainty arising from measurements.
 If all uncertainties (the intrinsic scatters and uncertainties arising from the correlations between
 the $L_{\rm P}$, $L_{\rm H\alpha}$, $L_{\rm 5100}$, and $L_{\rm bol}$) are combined together, the combined uncertainty increases to 0.36 dex.}

 We compared the Eddington ratios ($L_{\rm bol}$/$L_{\rm Edd}$, where $L_{\rm Edd}$ is the Eddington luminosity)
 of red quasars and unobscured type 1 quasars in Figure 12.
 The median Eddington ratios of unobscured type 1 quasars is 0.12 with an rms scatter of 0.34.
 Meanwhile, the measured Eddington ratios of three red quasars are 0.83, 0.30, and 0.52 for 0036$-$0113, 1209$-$0107, and 1307$+$2338, respectively.
 Among 213 unobscured type 1 quasars, only 55 unobscured type 1 quasars have higher Eddington ratios than 0.30.
 The probability is only 1.7\,\% that three randomly chosen unobscured type 1 quasars from the comparison sample
 have the Eddington ratios higher than 0.30.
 No systematic bias is expected between $L_{\rm bol}$/$L_{\rm Edd}$ from P$\beta$ (red quasars)
 and $L_{\rm 5100}$ with $\rm FWHM_{\rm H\beta}$ (unobscured type 1 quasars),
 except for an added scatter of 0.36 dex that comes from the scatter in the correlation of different quantities \citep{kim15b}.

 Since several works suggest that the Eddington ratios are dependent on $L_{\rm bol}$ (e.g., \citealt{lusso12,suh15}),
 we also show the Eddington ratio distribution of the sample in a narrow $L_{\rm bol}$ range
 that matches the $L_{\rm bol}$ of red quasars.
\ The objects 0036$-$0113 and 1307$+$2338 have similar $L_{\rm bol}$ of $\rm 10^{45.79}\,erg\,s^{-1}$ and $\rm 10^{45.50}\,erg\,s^{-1}$, respectively,
 but 1209$-$0107 has a much higher $L_{\rm bol}$ of $\rm 10^{46.51}\,erg\,s^{-1}$.
 For comparison, we selected 48 unobscured type 1 quasars with the same $L_{\rm bol}$ range
 ($10^{45.50} < L_{\rm bol} < 10^{45.79}\,{\rm erg\,s^{-1}}$) of the 2 red quasars 0036$-$0113 and 1307$+$2338.
 The median Eddington ratio of the 48 unobscured type 1 quasars changes to 0.15$\pm$0.33, but this is still much lower than those of the red quasars.
 Among the 48 unobscured type 1 quasars, only 9 unobscured type 1 quasars have higher Eddington ratios than
 the minimum Eddington ratio, 0.52, of 0036$-$0113 and 1307$+$2338.
 The probability is only 3.2\,\% that 2 randomly chosen unobscured type 1 quasars from the comparison sample
 have the Eddington ratios higher than 0.52.

 Previous studies \citep{urrutia12,kim15b} showed red quasars at different redshift
 have higher Eddington ratios than unobscured type 1 quasars.
 \cite{kim15b} measured Eddington ratios for 16 red quasars at $z \sim 0.7$
 and the 16 red quasars are included in our 20 red quasars.
 \cite{kim15b} compared the Eddington ratios of the 16 red quasars
 to those of unobscured type 1 quasars that are matched in $M_{\rm BH}$.
 We show how the Eddington ratios of 16 red quasars compare with those of $L_{\rm bol}$-matched unobscured type 1 quasars.
 The unobscured type 1 quasars are selected from the SDSS DR7 quasar catalog \citep{schneider10}
 with the same redshift range of the 16 red quasars ($0.56 < z < 0.84$).
 We note that the selection method of the comparison sample is identical to \cite{kim15b} and
 the details of the comparison sample selection are described in Section 2.1 of \cite{kim15b}.

 We divided the red quasars and the unobscured type 1 quasars into two luminosity bins.
 Figure 13 shows the comparison of the Eddington ratio distributions
 and our results show that the Eddington ratios of 16 red quasars at $z \sim 0.7$ are significantly higher (factors of $\sim$3 to 4)
 than those of unobscured type 1 quasars.
 Even if we assume a maximal amount of 35\,\% contamination on the Paschen line flux by a narrow line (see Section 3),
 the Eddington ratio decreases by only $\sim$15\,\% or 0.07\,dex, and we find that the analysis result does not change.

 For the low luminosity sample, 9 red quasars and 246 unobscured type 1 quasars are selected,
 and their $\log (M_{\rm BH}/M_{\rm \odot})$ values have ranges of 7.98--9.07 and 7.71--10.27
 for the red quasars and the unobscured type 1 quasars, respectively.
 The median Eddington ratios of the red quasars and the unobscured type 1 quasars are 0.62 and 0.15, respectively.
 The $D$ and $p$ values from a K-S test between these two distributions are 0.64 and $7.6 \times 10^{-4}$.
 The high luminosity sample includes 7 red quasars and 36 unobscured type 1 quasars.
 The $\log (M_{\rm BH}/M_{\rm \odot})$ values have ranges of 8.26--9.09 and 8.58--10.18
 for the red quasars and unobscured type 1 quasars, respectively.
 The median Eddington ratios of the red quasars and unobscured type 1 quasars are 0.97 and 0.35, respectively.
 The measured $D$ and $p$ values between these two distributions are 0.66 and $5.6 \times 10^{-3}$. 

 Overall, the additional analysis of the three red quasars and the re-analysis of the luminosity matched 16 red quasars in \citet{kim15b}
 strengthens the previous results of \citet{urrutia12} and \citet{kim15b}
 that red quasars tend to have higher Eddington ratios than unobscured type 1 quasars.
 Therefore, we suggest that many of the red quasars, if not all, are not seen as red simply because of the viewing angle (also see \citealt{onori17}).

\clearpage
\begin{sidewaystable*}[t]
\centering
\caption{Values of $M_{\rm BH}$, $L_{\rm bol}$, and $E(B-V)$ for red quasars \label{tbl2}}
\begin{tabular}{cccccccccc}
\hline\hline
\noalign{\smallskip}
Object name&    &       $M_{\rm BH}$&                                   $L_{\rm bol}$&                                  Eddington ratio&        $E(B-V)_{\rm cont}$\tablefootmark{a}&        $E(B-V)_{\rm cont}$\tablefootmark{b}&   $E(B-V)_{\rm line 1}$&       $E(B-V)_{\rm line 2}$&          $E(B-V)_{\rm BD}$\\
        &                       &       (${\rm 10^8}\,M_{\rm \odot}$)&  ($\rm 10^{46}\,erg\,s^{-1}$)& &       (mag)                                                           &       (mag)                           &                                               (mag)&                                  (mag)&                                  (mag)\\
\noalign{\smallskip}
\hline
\noalign{\smallskip}
0036$-$0113&    &       0.60$\pm$0.12&                  0.61$\pm$0.03&                  0.830&                  0.79&   0.96$\pm$0.10&  $>$1.58&                0.98$\pm$0.07&  2.54$\pm$0.26\tablefootmark{d}\\
0817$+$4354&    &       --&                                             --&                                             --&                             0.73&   --&                             --&                             2.40$\pm$0.19&  --\\
0825$+$4716&    &       12.25$\pm$2.98$\ast$&   10.40$\pm$1.02$\ast$&   0.693$\ast$&    0.71&   0.52$\pm$0.10&  1.59$\pm$0.27&  --&                             --\\
0911$+$0143&    &       2.59$\pm$0.44$\ast$&    1.54$\pm$0.16$\ast$&    0.486$\ast$&    --&             0.63$\pm$0.17&  --&                             --&                             --\\
0915$+$2418&    &       10.94$\pm$4.92$\ast$&   12.95$\pm$3.18$\ast$&   0.967$\ast$&    0.68&   0.36$\pm$0.12&  --&                             --&                             --\\
1113$+$1244&    &       3.09$\pm$0.52$\ast$&    8.92$\pm$0.67$\ast$&    2.360$\ast$&    1.15&   1.41$\pm$0.11&  0.45$\pm$0.09&  --&                             --\\
1209$-$0107&    &       8.75$\pm$2.71&                  3.25$\pm$0.21&                  0.303&                  0.89&   --&                             --&                             --&                             --\\
1227$+$5053&    &       2.87$\pm$1.00$\ast$&    2.26$\pm$0.49$\ast$&    0.644$\ast$&    0.38&   --&                             0.23$\pm$0.11&  --&                             -0.23$\pm$0.07\tablefootmark{c}\\
1248$+$0531&    &       2.49$\pm$0.32$\ast$&    1.93$\pm$0.12$\ast$&    0.633$\ast$&    --&             0.26$\pm$0.07&  --&                             --&                             --\\
1307$+$2338&    &       0.49$\pm$0.11&                  0.32$\pm$0.02&                  0.525&                  0.58&   0.75$\pm$0.09&  $>$1.47&                1.46$\pm$0.08&  1.74$\pm$0.10\tablefootmark{d}\\
1309$+$6042&    &       2.37$\pm$0.26$\ast$&    1.80$\pm$0.09$\ast$&    0.622$\ast$&    --&             0.95$\pm$0.13&  1.29$\pm$0.07&  --&                             --\\
1313$+$1453&    &       7.62$\pm$0.92$\ast$&    4.65$\pm$0.17$\ast$&    0.499$\ast$&    --&             1.12$\pm$0.15&  --&                             --&                             --\\
1434$+$0935&    &       0.95$\pm$0.11$\ast$&    3.88$\pm$0.13$\ast$&    3.350$\ast$&    --&             1.14$\pm$0.10&  0.81$\pm$0.08&  --&                             --\\
1532$+$2415&    &       6.12$\pm$4.87$\ast$&    2.24$\pm$1.05$\ast$&    0.299$\ast$&    0.70&   --&                             $>$2.11&                --&                             --\\
1540$+$4923&    &       11.86$\pm$7.13$\ast$&   2.73$\pm$1.11$\ast$&    0.188$\ast$&    --&             0.97$\pm$0.10&  --&                             --&                             --\\
1600$+$3522&    &       1.81$\pm$0.49$\ast$&    4.26$\pm$0.63$\ast$&    1.927$\ast$&    0.22&   --&                             --&                             --&                             --\\
1656$+$3821&    &       4.87$\pm$0.98$\ast$&    4.66$\pm$0.53$\ast$&    0.783$\ast$&    0.61&   0.88$\pm$0.16&  $>$0.30&                --&                             --\\
1720$+$6156&    &       1.31$\pm$0.33$\ast$&    2.58$\pm$0.38$\ast$&    1.613$\ast$&    --&             1.50$\pm$0.09&  0.57$\pm$0.28&  --&                             --\\
2325$-$1052&    &       2.71$\pm$1.59$\ast$&    2.02$\pm$0.73$\ast$&    0.610$\ast$&    0.50&   --&                             0.64$\pm$0.21&  --&                             0.56$\pm$0.01\tablefootmark{d}\\
2339$-$0912&    &       5.95$\pm$0.95$\ast$&    8.03$\pm$0.55$\ast$&    1.102$\ast$&    1.06&   1.25$\pm$0.06&  0.53$\pm$0.07&  --&                             --\\
\noalign{\smallskip}
\hline
\end{tabular}
\tablefoot{\\
\tablefoottext{$\ast$}{$M_{\rm BH}$, $L_{\rm bol}$ values, and Eddington ratios are adopted from \cite{kim15b}.}\\
%$\ast\ast$: The $E(B-V)_{\rm cont}$ values are adopted from \cite{glikman07}, which uncertainties are not available.\\
\tablefoottext{a}{$E(B-V)_{\rm cont}$ values are adopted from \cite{glikman07}, which uncertainties are not available.}\\
\tablefoottext{b}{$E(B-V)_{\rm cont}$ values are adopted from \cite{urrutia09}.}\\
\tablefoottext{c}{$E(B-V)_{\rm BD}$ values are measured with total line luminosities.}\\
\tablefoottext{d}{$E(B-V)_{\rm BD}$ values are measured with broad line luminosities.}
}
\end{sidewaystable*}

\section{Conclusions}
 So far, several explanations for the red colors of red quasars have been suggested; for example,
 (i) there is dust in their host galaxy, (ii) these quasars are intrinsically red,
 (iii) the redness is attributable to high $\rm CF_{HD}$, and (iv) the red color can be attributed to a moderate viewing angle.
 In order to investigate the origin of the red colors of red quasars,
 we used 20 red quasars at $z \sim$ 0.3 and 0.7 in this study.

 We compared the luminosity ratios from hydrogen Balmer to Paschen lines of red quasars to those of unobscured type 1 quasars adopted from \citet{kim10}.
 We find that the line luminosity ratios of red quasars are almost six times higher than those of unobscured type 1 quasars,
 and the extinction correction based on $E(B-V)$ estimated from the continuum shape
 can bring the similar mean P$\beta$/H$\beta$ luminosity ratio to that of unobscured type 1 quasars (albeit with a large scatter).
 This result suggests that dust extinction is responsible for the red colors of red quasars.

 Moreover, we examined if the characteristics of red quasars can be explained by other reasons.
 We examined whether the unusual line luminosity ratios of red quasars can be explained without the dust extinction effects.
 We compared the observed line luminosity ratios of red quasars to the theoretically expected line luminosity ratios computed from \texttt{CLOUDY} code.
 However, the observed line luminosity ratios of $\sim 63$\,\% (5 out of 8) red quasars cannot be explained by any physical conditions of BELRs.
 We also examined whether red quasars have unusually higher $\rm{CF_{HD}}$ than unobscured type 1 quasars,
 but we do not find clear evidence of NIR excess for a given Paschen line luminosity.
 Finally, using a dust-insensitive diagnostic P$\beta$ line,
 we find that 19 red quasars -- three at $z \sim 0.3$ (new measurements) and 16 at $z \sim 0.7$ \citep{kim15b} --
 have Eddington ratios that are significantly ($\sim5$\,times) higher than
 those of unobscured type 1 quasars.
 This result disfavors the scenario in which the red colors of red quasars are caused by
 a moderate viewing angle that passes though dust torus.

 Our results and previous results from studies of red quasars have found that these quasars have
 (i) high luminosity ratios from Balmer to Paschen lines,
 (ii) high BH accretion rates \citep{kim15b},
 (iii) enhanced star formation activities \citep{georgakakis09},
 (iv) a high fraction of merging features \citep{urrutia09,glikman15},
 and (v) young radio jets \citep{georgakakis12}. Based on these findings, we conclude that red quasars are dust-extincted quasars in the intermediate stage galaxies between ULIRGs and unobscured type 1 quasars
 and the dust extinction is caused by the remaining dust in their host galaxies.

\begin{acknowledgements}

 This work was supported by the Creative Initiative Program of the National Research Foundation of Korea (NRF),
 No. 2017R1A3A3001362, funded by the Korea government.
 D.K. acknowledges support by the National Research Foundation of Korea to
 the Fostering Core Leaders of the Future Basic Science Program, No. 2017-002533.\\

 M.I. and D.K. are Visiting Astronomers at the Infrared Telescope Facility,
 which is operated by the University of Hawaii under Cooperative Agreement no. NNX-08AE38A
 with the National Aeronautics and Space Administration, Science Mission Directorate, Planetary Astronomy Program.\\
%http://irtfweb.ifa.hawaii.edu/research/acknowledge.php

 We thank Eilat Glikman for sharing spectra of the red quasars listed in \cite{glikman07}.
\end{acknowledgements}

\bibliographystyle{aa} % style aa.bst
\bibliography{redcolor} % your references Yourfile.bib

\clearpage
%%%%%%%%%%%%%%%%%%%%<<<<<<              FIGURE                  >>>>>%%%%%%%%%%%%%%%%%%%%%%%%%%%%%%%%%

%%%%%%%%%%%%%%%%%%%%<<<<<<              TABLE                   >>>>>%%%%%%%%%%%%%%%%%%%%%%%%%%%%%%%%%

%\appendix
\begin{appendix}
\section{Near-infrared spectra of red quasars}
 We provide 0.8--2.5\,$\mu$m spectra of two red quasars (0036$-$0113 and 1307$+$2338) taken with NASA IRTF.
 Table 3 is an example spectrum of 0036$-$0113, and the full version of the spectra
 is available in machine readable table form.

%\centering
\begin{table}
\centering
\caption[]{Spectrum of 0036$-$0113\label{tbl3}}
\begin{tabular}{ccc}
\hline\hline
\noalign{\smallskip}
$\lambda$&              $f_{\lambda}$&                                                  $f_{\lambda}$ Uncertainty\\
($\rm \AA{}$)&  ($\rm{erg~s^{-1}~cm^{-2}~\AA{}^{-1}}$)& ($\rm{erg~s^{-1}~cm^{-2}~\AA{}^{-1}}$)\\
\noalign{\smallskip}
\hline
\noalign{\smallskip}
10900&  4.470E-18&  2.652E-17\\
10903&  5.215E-17&  2.763E-17\\
10905&  4.738E-17&  2.575E-17\\
10908&  5.146E-17&  2.566E-17\\
10911&  7.565E-17&  2.431E-17\\
10913&  6.210E-17&  2.379E-17\\
10916&  1.274E-17&  2.386E-17\\
10919&  2.437E-17&  2.621E-17\\
10921&  8.003E-17&  2.719E-17\\
10924&  8.594E-17&  2.935E-17\\
\noalign{\smallskip}
\hline
\end{tabular}
\tablefoot{This table lists only a part of the spectrum of 0036$-$0113. The entire spectra of two red quasars
(0036$-$0113 and 1307$+$2338) are available in ascii format from the electronic version of the Journal.}
\end{table}
\end{appendix}

\end{document}